\documentclass[preprint,superscriptaddress,amsmath,amssymb,aps,eqsecnum,nofootinbib]{revtex4}
\usepackage[dvipdfmx]{graphicx}

\usepackage{subfigure}
\usepackage{bm}
\usepackage{mathrsfs}

\begin{document}
\title{Vanishing Love of Black Holes in General Relativity:\\
From Spacetime Conformal Symmetry of a Two-dimensional Reduced Geometry}
\author{Takuya Katagiri}
\affiliation{Astronomical Institute, Graduate School of Science, Tohoku University, Aoba, Sendai 980-8578, Japan}
\affiliation{Niels Bohr International Academy, Niels Bohr Institute, Blegdamsvej 17, 2100 Copenhagen, Denmark}
\author{Masashi Kimura}
\affiliation{Department of Informatics and Electronics, Daiichi Institute of Technology, Tokyo 110-0005, Japan}
\affiliation{Department of Physics, Rikkyo University, Toshima, Tokyo 171-8501, Japan}
\author{Hiroyuki Nakano}
\affiliation{Faculty of Law, Ryukoku University, Kyoto 612-8577, Japan}
\author{Kazuyuki Omukai}
\affiliation{Astronomical Institute, Graduate School of Science, Tohoku University, Aoba, Sendai 980-8578, Japan}
\date{\today}

\begin{abstract}
We study the underlying structure of the vanishing of the Love numbers of both Schwarzschild and Kerr black holes in terms of spacetime conformal symmetry in a unified manner for the static spin-$s$ fields. The perturbations can be reduced with the harmonic decomposition to a set of infinite static scalar fields in a two-dimensional anti-de Sitter spacetime~$({\rm AdS}_2)$. In the reduced system, each scalar field is paired with another, implying that all multipole modes of the perturbation can be regarded as symmetric partners, which can be understood from the property of the supersymmetry algebra. The generator of the supersymmetric structure is constructed from a closed conformal Killing vector field of ${\rm AdS}_2$. 
The associated conserved quantity allows one to show no static 
response, i.e., vanishing of the Love and dissipation numbers. We also discuss the vanishing Love numbers of the Kerr black hole with the nonzero dissipation numbers for the non-axisymmetric perturbations in terms of a radial constant found in a parallel manner as the axisymmetric field case even though the interpretation for the structure is controversial. The symmetric structure corresponds to the ``ladder'' symmetry in~Hui {\it et al.}~[JCAP \textbf{01}, no.01, 032 (2022)] although the geometrical origin is different. Our ladder operator includes the generators of hidden symmetries in previous works.

\end{abstract}
\maketitle

\section{Introduction}
The black hole is one of the most surprising predictions of General Relativity. Its only component is the concept of spacetime. 
Black holes have remarkable properties in General Relativity, i.e., no hair~\cite{PhysRevLett.26.331,Bekenstein:1972ky} and uniqueness~\cite{Hawking:1971vc,Robinson:1975bv,PhysRevD.81.024033}. 
As a consequence, all the astrophysical black holes are believed to be well described by the Kerr geometry, which is characterized solely by the mass and spin parameters. 
On the other hand, General Relativity might need modification in the vicinity of the event horizon.
In modified theories of gravity, black holes can possess a scalar hair~\cite{PhysRevLett.106.151104,PhysRevD.93.044047}, unlike the Kerr black holes. 
The detection of the deviation from the Kerr geometry by future gravitational-wave observations can be a smoking gun for such new physics.

Binary black hole systems are strong candidates for gravitational-wave sources~\cite{LIGOScientific:2016aoc} and also work as astrophysical laboratories to test strong gravitational fields~\cite{Berti:2015itd,Berti:2005ys}. 
During the early inspiral phase of a compact binary, the two bodies behave as point masses, as their internal structure does not affect the orbital motion~\cite{PhysRev.136.B1224}. 
As the orbital separation sufficiently decreases due to the gravitational-wave emission, however, the tidal interaction between the bodies becomes remarkable and higher-order post-Newtonian correction comes into effect. 
Its correction appears at the $5$PN order in the gravitational waveform as a function of a quantity called {\it tidal deformability}, which characterizes how the bodies deform as a static response to the perturbative tidal field generated by each other~\cite{Vines:2011ud}. 
The tidal deformability is encoded in a set of the tidal Love numbers, which are determined by the internal structure of the body~\cite{Hinderer:2007mb,Binnington:2009bb,Damour:2009vw,poisson_will_2014}. 
This means that a measurement of the tidal Love numbers from a gravitational-wave signal provides information about the internal structure of the bodies as an inverse problem~\cite{Flanagan:2007ix}. Indeed, the equation of state of a neutron star has been constrained from GW170817~\cite{Flanagan:2007ix,LIGOScientific:2017vwq,LIGOScientific:2017ync}. In the same manner, it is expected that a test of the strong-field gravity can be performed with tidal Love numbers~\cite{Cardoso:2017cfl}. 

An intriguing result regarding the tidal Love numbers is their vanishing for (four-dimensional) black holes in vacuum in General Relativity, i.e., Schwarzschild~\cite{Binnington:2009bb,Gralla:2017djj} and Kerr black holes~\cite{LeTiec:2020spy,Chia:2020yla,LeTiec:2020bos,Charalambous:2021mea}. 
The Schwarzschild and Kerr black holes either have the vanishing Love numbers for spin-$0$ or spin-$1$ field perturbations~\cite{Hui:2020xxx,Charalambous:2021mea}. 
On the other hand, in some modified theories of gravity, black holes can have nonzero tidal Love numbers~\cite{Cardoso:2017cfl,Cardoso:2018ptl}. 
One might then think that an additional field can endow a black hole with nonzero Love numbers. 
In fact, the Love numbers of black holes can be nonzero in cases with an anisotropic fluid~\cite{Cardoso:2019upw,Cardoso:2021wlq} or with a bosonic scalar-field condensate~\cite{DeLuca:2021ite} within General Relativity. 
In addition, black holes have zero Love numbers
in some modified theories of gravity or in the presence of some kind of ``matter'' field, e.g., the Schwarzschild black hole in the Brans-Dicke theory and the Reissner-Nordstr\"{o}m black hole in the Einstein-Maxwell theory~\cite{Cardoso:2017cfl,Cardoso:2019upw}. 
Similarly, even in vacuum in General Relativity, the Schwarzschild-Tangherlini spacetime has nonzero Love numbers~\cite{Hui:2020xxx}. 
In other contexts, several literatures~\cite{Porto:2016zng,Penna:2018gfx,Hui:2021vcv} pointed out that from the viewpoint of an effective field theory approach~\cite{Goldberger:2004jt,Goldberger:2005cd,Kol:2011vg}, the no tidal response of the Schwarzschild black hole is puzzling and seems to be a result of fine-tuning. These suggest that some non-trivial underlying structure is prohibiting finite Love numbers.

A linear response of compact objects is also characterized by other quantities which are imprinted in gravitational waveforms coming from an inspiraling binary, i.e., dissipation numbers that quantify the dissipation of an external tidal field. For a time-varying external field, black holes have nonzero dissipation numbers due to the presence of the event horizon, while horizonless compact objects have their vanishing. This property may allow for constraining a quantum correction at the horizon scale with future gravitational-wave observation~\cite{PhysRevLett.122.081301,Cardoso:2019rvt}. For a static external field, Schwarzschild black holes have vanishing Love and dissipation numbers, meaning no static response. Kerr black holes have non-zero dissipation numbers for non-axisymmetric static fields because of a relative motion arising from rotation, while both the Love and dissipation numbers vanish for axisymmetric fields, i.e., no static response.

Several authors independently argue that the vanishing of Love numbers may be a result of ``hidden'' symmetries governing linear perturbations.\footnote{Here, the terminology~``hidden'' symmetries means that they do not correspond to isometries of a spacetime.} Penna~\cite{Penna:2018gfx} suggested in the context of the black-hole membrane paradigm that the emergent local Carroll symmetry~\cite{Duval:2014lpa} in the near-horizon region may have a role in explaining the no static response of the Schwarzschild black hole. 
Charalambous, Dubovsky, and Ivanov~\cite{Charalambous:2021kcz} argued that a long-wavelength spin-$s$ field perturbation in the Kerr spacetime has a hidden ${\rm SL}(2,\mathbb{R})\times U(1)$ symmetry dubbed Love symmetry and spin-$s$-field Love numbers of the Kerr black hole vanish as a consequence of this symmetry based on the ${\rm SL}(2,\mathbb{R})$ representation theory. 
Recently, Hui, Joyce, Penco, Santoni, and Solomon~\cite{Hui:2021vcv} showed the vanishing of scalar-field Love numbers from the perspective of two types of hidden symmetries, one of which originates from the presence of a Killing vector field in a three-dimensional Euclidian anti-de Sitter space~(see also Ref.~\cite{BenAchour:2022uqo}). 
More recently, Charalambous, Dubovsky, and Ivanov~\cite{Charalambous:2022rre} conducted a detailed study on the vanishing of spin-$s$-field Love numbers of Kerr-Newman black holes in terms of ${\rm SL}(2,\mathbb{R})$ representations.

The recent arguments above on the connection between the static response and a hidden symmetry are reminiscent of the Laplace-Runge-Lenz vector, the conserved quantity associated with a dynamical symmetry in the Kepler problem~\cite{goldstein:mechanics}. The Laplace-Runge-Lenz vector explains the absence of the periapsis shift in the inverse-square central force.
The Laplace-Runge-Lenz vector also appears in the structure of the hydrogen atom and explains the degeneracy of the energy levels with different orbital angular momenta~\cite{2003AmJPh..71..171V}. 
Some hidden symmetries may also explain the vanishing of the Love numbers of black holes based on the analogy of quantum mechanics and the black-hole perturbation theory as in Refs.~\cite{PhysRevLett.52.1361,PhysRevD.30.295,PhysRevD.43.605,PhysRevD.101.024008,Cardoso:2017qmj,Cardoso:2017egd,PhysRevD.106.044052,Bonelli:2021uvf}. However, we still have a fundamental question: {\it why does such a hidden symmetric structure exist?} 

It is natural to expect that a geometrical property such as spacetime symmetry gives symmetry for perturbation fields in a given spacetime. In addition, if some hidden symmetric structure provides a useful way to understand the vanishing of Love numbers as several authors argued, the strategy should work for spin-s fields in a unified manner because the vanishing Love number is a common property of all the scalar, vector, and tensor-field perturbations to Schwarzschild and Kerr black holes. In this paper, for the above problems, we study the underlying symmetric structure of the vanishing of the spin-$s$-field Love numbers of the Schwarzschild and Kerr black holes in terms of spacetime conformal symmetry in a unified manner. We reduce a perturbation with the harmonic decomposition into a set of infinite static scalar fields in the two-dimensional anti-de Sitter spacetime~$({\rm AdS}_2)$. We then discuss a symmetric structure generated by a ladder operator constructed from a closed conformal Killing vector field of ${\rm AdS}_2$.\footnote{The hidden symmetric structure corresponds to the ``ladder'' symmetry in Ref.~\cite{Hui:2021vcv} although the geometrical origin is different. Hui~{\it et al}.~\cite{Hui:2021vcv} gave a generator, i.e., ladder operator, arising from a Killing vector field of a Euclidean ${\rm AdS}_3$ for a scalar field case~$s=0$; they also found a ladder operator for vector~$(s=1)$ and tidal field~$(s=2)$ cases but its construction seems to be heuristic. Our viewpoint gives a ladder operator for spin-$s$ fields from a closed conformal Killing vector field in ${\rm AdS}_2$ in a unified manner. Our operator includes the ladder operators in Ref.~\cite{Hui:2021vcv}; the ladder operators in Ref.~\cite{BenAchour:2022uqo} and the mass ladder operator in Refs.~\cite{Cardoso:2017qmj,Cardoso:2017egd,PhysRevD.106.044052,Katagiri:2021scx,Katagiri:2021xig} for a scalar field case.}  The no static response of the Schwarzschild black hole can be understood in terms of the associated conserved quantity. This is also the case of the Kerr black hole for axisymmetric perturbation fields. We further discuss the vanishing Love numbers of the Kerr black hole with the nonzero dissipation numbers for non-axisymmetric perturbations in terms of a radial constant which is found in a parallel manner as the axisymmetric field case even though the interpretation for the aforementioned structure is controversial~\cite{Hui:2021vcv, Hui:2022vbh,Charalambous:2022rre}. To our knowledge, this is the first attempt to explain no static response of the Schwarzschild and Kerr black holes for spin-$s$ fields in a unified manner based on a symmetric approach from a geometrical point of view.\footnote{When the manuscript had almost completed, Ref.~\cite{Charalambous:2022rre} appeared. The generator in Ref.~\cite{Charalambous:2022rre} corresponds to the covariant Lie derivatives with respect to the tetrad transformations of the Newman-Penrose approach. It appears to come that this and ours are one of several aspects of the vanishing of Love numbers from different perspectives. We mention here that for a scalar field case, the generators of ${\rm SL}(2,\mathbb{R})$ symmetry in Refs.~\cite{Charalambous:2021kcz,Charalambous:2022rre} are Killing vector fields of ${\rm AdS}_2$. }

The organization of the rest is as follows. Section~\ref{Sec:reviewofstaticrespose} presents a brief review of a static response of the Schwarzschild black hole to spin-$s$ field perturbations. In Sec.~\ref{Sec:HiddenSymmetryfromAdS2}, we reduce the problem of the static spin-$s$ field perturbation on the Schwarzschild black hole to that of static scalar fields in ${\rm AdS}_2$, and further show a hidden symmetric structure governing the perturbations from the perspective of spacetime conformal symmetry of the reduced geometry. We then explain in Sec.~\ref{Sec:VanishingLovefromSymmetry} the no static response of the Schwarzschild black hole in terms of the symmetric structure. In Sec.~\ref{Sec:VanishingLoveofKerrBH}, we discuss the relation between the vanishing Love numbers and a hidden symmetric structure. 
Section~\ref{Sec:Conclusion} is devoted to the summary and implication of this paper. Appendices give the supplementary materials; a slowly-varying perturbation case, derivation of a ladder operator from spacetime conformal symmetry and the application of the operator to perturbations to the Schwarzschild-Tangherlini black hole~\cite{Tangherlini:1963bw}.

\section{Static response of Schwarzschild black holes}
\label{Sec:reviewofstaticrespose}
In this section, we briefly review a linear static response of Schwarzschild black holes to  external spin-$s$ fields, i.e., scalar~($s=0$), vector~($s=1$), and gravitational~($s=2$) fields~\cite{Binnington:2009bb}. We assume that the amplitude of the field is small, and its wavelength is much longer than the horizon scale. The field is then described by a linear static perturbation theory of the Schwarzschild black hole.

\subsection{Linear perturbation theory of the Schwarzschild black hole}
We here review the linear perturbation theory to the Schwarzschild black hole for the tidal field~$(s=2)$ and its generalization for the spin-$s$ fields. The Schwarzschild black hole spacetime is described by
\begin{equation}
\label{Schwarzschildmetric}
g_{\mu\nu}^{(0)}dx^\mu dx^\nu=-\frac{\Delta}{z^2}dt^2+\frac{z^2}{\Delta}dz^2+z^2d\Omega^2,
\end{equation}
where
\begin{equation}
\Delta=z(z-1),
\end{equation}
and $d\Omega^2=d\theta^2+\sin^2\theta d\varphi^2$ is the line element of a two-dimensional unit sphere~$S^2$. In the current coordinate system~$(t,z,\theta,\varphi)$, the radial coordinate~$z$ is defined in the range $z\in (1,\infty)$, where $z=1$ and $\infty$ correspond to the locations of the event horizon and spatial infinity, respectively. 

We first explain the linear theory of a tidal field perturbation, i.e., $s=2$. On the Schwarzschild background~\eqref{Schwarzschildmetric}, the linearly perturbed metric takes a form,
\begin{equation}
g_{\mu \nu}=g_{\mu \nu}^{(0)}+h_{\mu \nu},
\end{equation}
where $h_{\mu \nu}$ is a linear perturbation. Each independent component of $h_{\mu \nu}$ is classified as either scalar or vector part on $S^2$, which can be expanded in terms of scalar spherical harmonics or vector spherical harmonics, respectively~\cite{Higuchi:1986wu,Kodama:2003jz}. The scalar and vector parts correspond to even and odd parts, respectively, under the parity transformation~$(\theta,\varphi)\to(\pi-\theta, \varphi+\pi)$. The former and latter are also called the polar-type and axial-type perturbations, respectively. Thus, we have
\begin{equation}
h_{\mu \nu}=h_{\mu \nu}^{\rm (polar)}+h_{\mu \nu}^{\rm (axial)},
\end{equation}
where $h_{\mu \nu}^{\rm (polar)}$ and $h_{\mu \nu}^{\rm (axial)}$ denote the polar- and axial-type perturbations, respectively. 

In the Regge-Wheeler gauge~\cite{Regge:1957td}, the linearized Einstein equation for the axial- and polar-type perturbations is reduced to two independent master equations. One for the axial-type perturbation~$\Phi_\ell^-$ is the so-called Regge-Wheeler equation~\cite{Regge:1957td} in the static limit:
\begin{equation}
\label{staticRWeq:s=2}
\frac{\Delta}{z^2}\frac{d}{dz}\left(\frac{\Delta}{z^2}\frac{d\Phi_\ell^-}{dz}\right)-\frac{\Delta}{z^2}\left[\frac{\ell(\ell+1)}{z^2}-\frac{3}{z^3}\right]\Phi_\ell^-=0,
\end{equation}
where $\ell=2,3,\cdots$ is the index of multipoles. Note that the azimuthal number~$m$ does not appear because of the spherical symmetry of the background spacetime. The other for the polar-type perturbation is the so-called Zerilli equation~\cite{PhysRevD.2.2141}. As shown in Appendix~\ref{Appendix:symmetryofpolarperturbation}, with the Chandrasekhar transformation~\cite{Chandrasekhar:1975zza}, the solutions of the static Zerilli equation can be generated from that of the static Regge-Wheeler equation~\eqref{staticRWeq:s=2}.\footnote{The master variable of the static polar-type perturbation can be defined well in the static limit as well~\cite{Jhingan:2002kb}.} 

We next give a perturbation equation for the spin-$s$ fields. Static scalar field~$(s=0)$, vector field~$(s=1)$, and axial-type tidal field~$(s=2)$ perturbations are uniformly described by~\cite{Kodama:2003jz,Berti:2009kk,Hatsuda:2021gtn}
\begin{equation}
\label{staticRWeq:review}
\frac{\Delta}{z^2}\frac{d}{dz}\left(\frac{\Delta}{z^2}\frac{d\Phi_\ell^{(s)}}{dz}\right)-\frac{\Delta}{z^2}\left[\frac{\ell(\ell+1)}{z^2}-\frac{s^2-1}{z^3}\right]\Phi_\ell^{(s)}=0,
\end{equation}
where $\ell\ge s$. In this work, we regard Eq.~\eqref{staticRWeq:review} as a fundamental equation governing static spin-$s$ field perturbations to the Schwarzschild black hole in a unified manner because the polar-type tidal field perturbation can also be described by the axial-type one with the Chandrasekhar transformation~\cite{Chandrasekhar:1975zza} as stated above.

\subsection{Spin-$s$-field Love and dissipation numbers}
Here, we introduce the notion of the spin-$s$-field Love and dissipation numbers. We consider asymptotic behaviors of solutions of Eq.~\eqref{staticRWeq:review} at the horizon and large distances. On the one hand, at the horizon~$z=1$, we have two linearly independent asymptotic solutions, i.e., 
\begin{equation}
\label{asymptoticsolatBH}
\Phi_\ell^{(s)}|_{z\to 1}\sim{\rm const.}, ~~\ln(1-1/z).
\end{equation}
Note that, at the horizon, the former and latter are regular and logarithmically divergent, respectively. On the other hand, at large distances~$z\gg1$, we have two linearly independent asymptotic solutions, i.e., 
\begin{equation}
\label{asymptoticsolatinfinity}
\Phi_\ell^{(s)}|_{z\gg1}\sim z^{\ell+1},~~z^{-\ell}.
\end{equation} 
A generic expectation might be that in the distant region~$z\gg 1$, the regular solution at the horizon, i.e., 
$\Phi_\ell^{(s)}|_{z\to 1}\sim{\rm const.}$
would be written in the form of a linear combination of two linearly independent solutions, i.e.,
\begin{equation}
\Phi_\ell^{(s)}|_{z\gg 1}\propto z^{\ell+1}\left[1+\mathcal{O}(1/z)+2k_\ell^{(s)} z^{-2\ell-1}\left(1+\mathcal{O}(1/z)\right)\right].
\end{equation} 
Physically, the first and second terms are interpreted as the external perturbation field and the static response of the black hole, respectively. The coefficient~$k_\ell^{(s)}$ is generically complex. The real part is called {\it spin-$s$-field Love numbers}~\cite{Cardoso:2017cfl,DeLuca:2021ite}. {\it Dissipation numbers} can be read off from the imaginary part~\cite{Chia:2020yla,Charalambous:2021mea}. In the same manner, the tidal Love and dissipation numbers for the polar-type gravitational perturbation can also be defined~\cite{poisson_will_2014,Binnington:2009bb}.

\subsection{Example: lowest multipole~$(\ell=s)$ case}
As an example, we show no static response to the lowest multipole mode of the physical solution. The analytic expressions for $\Phi_s^{(s)}$ are given, respectively, by
\begin{equation}
\Phi_0^{(0)}=C_1^{(0)}z+C_2^{(0)}z \ln \left(1-1/z\right),
\end{equation}
for $s=0$, and
\begin{equation}
\Phi_1^{(1)}=C_1^{(1)}z^2+C_2^{(1)}\left[1+2z+2z^2 \ln \left(1-1/z\right)\right],
\end{equation}
for $s=1$, and
\begin{equation}
\begin{split}
\label{generalsol:l=2}
\Phi_2^{(2)}= &C_{1}^{(2)}z^3+C_{2}^{(2)}z^{-1}\left[3+4z+6 z^2+12 z^3+12z^4\ln\left(1-1/z\right)\right],
\end{split}
\end{equation}
for $s=2$, where $C_1^{(s)}$ and $C_2^{(s)}$ are constants. Note here that the term of $C_{2}^{(s)}$ has a logarithmic contribution, which is divergent in the horizon limit~$z\to 1$. To ensure the regularity of the perturbations at the horizon, we have to set $C_{2}^{(s)}=0$. Therefore, the solution regular at the horizon takes the form, 
\begin{align}
\label{regularsol:l=s}
\Phi_s^{(s)}= &C_{1}^{(s)}z^{s+1}.
\end{align}
Note that this solution has no decaying term of $z$ at large distances~$z\gg1$. Thus, the coefficient~$k_s^{(s)}$ is precisely zero for the Schwarzschild black hole, meaning no static response. One can show that the higher-multipolar spin-$s$-field Love and dissipation numbers also vanish in the same manner. The tidal Love and dissipation numbers for the polar-type tidal field perturbation are also precisely zero as shown in Appendix~\ref{Appendix:symmetryofpolarperturbation}.

\section{Hidden supersymmetric structure of static perturbations}
\label{Sec:HiddenSymmetryfromAdS2}
In this section, we show the existence of a ``supersymmetric'' structure governing static spin-$s$ field perturbations to the Schwarzschild black hole from a geometrical point of view. We first demonstrate that the static spin-$s$ field perturbation to the Schwarzschild black hole can be reduced to a set of infinite static scalar fields in the two-dimensional anti-de Sitter spacetime~(${\rm AdS}_2$). We then construct a ladder operator that is a generator of the symmetric structure, from a particular spacetime conformal symmetry of ${\rm AdS}_2$, and further derive the conserved quantity associated with the symmetric structure. In Appendix~\ref{Appendix:Reductionofalongwavelengthperturbation}, we show that the following discussion is completely parallel also for a slowly-varying perturbation.

\subsection{Reduction to ${\rm AdS}_2$}
The static spin-$s$ field perturbation satisfies
\begin{equation}
\label{staticRWeq}
\frac{\Delta}{z^2}\frac{d}{dz}\left(\frac{\Delta}{z^2}\frac{d\Phi_\ell}{dz}\right)-\frac{\Delta}{z^2}\left[\frac{\ell(\ell+1)}{z^2}-\frac{s^2-1}{z^3}\right]\Phi_\ell=0,
\end{equation}
which is the same as Eq.~\eqref{staticRWeq:review}. Henceforth, we omit the superscript~$s$ denoting the spin of the field. We rewrite Eq.~\eqref{staticRWeq} into more convenient form for later discussion:
\begin{align}
\label{NHstaticZeq}
\frac{d}{dz}\left(\Delta\frac{d\phi_\ell}{dz}\right)-\left(\ell(\ell+1)-\frac{s^2}{z}\right)\phi_\ell=&0,
\end{align}
where we have defined
\begin{equation}
\label{fieldredefinition}
\phi_\ell(z)=\frac{\Phi_\ell}{z}.
\end{equation}

In what follows, we shall see that Eq.~\eqref{NHstaticZeq} can be identified as the equation of motion for a static scalar field in ${\rm AdS}_2$. Consider a line element in the static coordinates~$(t,z)$,
\begin{equation}
\label{AdS2metric}
g_{ab}^{\rm AdS_2} dx^a dx^b
=-\Delta dt^2+\frac{1}{\Delta}dz^2,
\end{equation}
where $a, b$ run $t$ and $z$. The scalar curvature of this metric is calculated as
\begin{equation}
R^{(2)}=-2.
\end{equation}
The spacetime described by Eq.~\eqref{AdS2metric} has a negative constant curvature, i.e., ${\rm AdS}_2$. The d'Alembertian in ${\rm AdS}_2$ is given by
\begin{equation}
\label{BoxAdS2general}
\square_{{\rm AdS}_2}:=-\frac{1}{\Delta}\partial_t^2+\partial_z\left(\Delta\partial_z\right).
\end{equation}
It should be noted that the first term of the left-hand side of Eq.~\eqref{NHstaticZeq} corresponds to the d'Alembertian~\eqref{BoxAdS2general} for a static field, i.e.,
\begin{equation}
\label{BoxAdS2}
\square_{{\rm AdS}_2}=\frac{d}{dz}\left(\Delta\frac{d}{dz}\right).
\end{equation}
This indicates that the variable~$\phi_\ell$ in Eq.~\eqref{NHstaticZeq} can be identified as a scalar field in ${\rm AdS}_2$ of Eq.~\eqref{AdS2metric} satisfying an equation of motion,
\begin{align}
\label{StaticscalarAdS2}
\left[\square_{{\rm AdS}_2}-\left(\ell(\ell+1)-\frac{s^2}{z}\right)\right]\phi_\ell=&0.
\end{align}
Thus, the static spin-$s$ field perturbation to the Schwarzschild black hole can be reduced to a set of infinite static scalar fields in ${\rm AdS}_2$ of Eq.~\eqref{AdS2metric}. Note that this reduction differs from the so-called near-horizon limit for extremal black holes~\cite{Bardeen:1999px} because the emergence of the ${\rm AdS}_2$ geometry is not inherited from the enhancement of the isometry of the background spacetime itself~\cite{Castro:2010fd,Bertini:2011ga}.

We comment that the reduction above corresponds to the static limit of that of a slowly varying time-dependent perturbation. As shown in Appendix~\ref{Appendix:Reductionofalongwavelengthperturbation}, a slowly varying time-dependent spin-$s$ field can be reduced to a time-dependent scalar field in ${\rm AdS}_2$. The equation of motion for that takes the same form as Eq.~\eqref{StaticscalarAdS2}. The following argument works even for the case of the time-dependent field in an almost parallel manner.

\subsection{Ladder operator from spacetime conformal symmetry}
We define a generator of a hidden symmetric structure, i.e., ladder operator, based on the ${\rm AdS}_2$ geometry,
\begin{equation}
\label{LOs}
\mathcal{D}_{k_\pm}:=\mathcal{L}_\zeta-\frac{k_\pm}{2}\left(\nabla_a\zeta^a-\frac{s^2}{k_\pm^2}\right),
\end{equation}
where
\begin{equation}
\label{kpm}
k_+=-\ell-1,~~k_-=\ell.
\end{equation}
Here, $\mathcal{L}_\zeta$ is the Lie derivative with respect to a vector field~$\zeta^a$ called a closed conformal Killing vector field in ${{\rm AdS}_2}$ of Eq.~\eqref{AdS2metric},
\begin{equation}
\label{CCKV}
\zeta^a \frac{\partial}{\partial x^a} =\Delta \frac{\partial}{\partial z}.
\end{equation}
which satisfies a conformal Killing equation,
\begin{equation}
\mathcal{L}_\zeta  g_{ab}=\left(\nabla_c\zeta^c\right)g_{ab},
\end{equation}
with the closed condition,
\begin{equation}
\nabla_a\zeta_b=\nabla_b\zeta_a.
\end{equation}
The operator~$\mathcal{D}_{k_\pm}$ in Eq.~\eqref{LOs} can be written as 
\begin{equation}
\label{laderoperator}
\mathcal{D}_{k_\pm}=\Delta\frac{d}{dz}-\frac{k_\pm}{2}\left(2z-1-\frac{s^2}{k_\pm^2}\right).
\end{equation}
We here point out that the ladder operator satisfies a commutation relation,
\begin{equation}
\label{commutationrelation}
\left[\Box_{{\rm AdS}_2}+\frac{s^2}{z},\mathcal{D}_{k_\pm}\right]=-2k_\pm\mathcal{D}_{k_\pm}+\left(\nabla_a\zeta^a\right)\left[\Box_{{\rm AdS}_2}-\left(k_\pm(k_\pm+1)-\frac{s^2}{z}\right)\right].
\end{equation}
Note that the construction of the ladder operator is coordinate-independent as seen in Appendix~\ref{Appendix:ConditionforVerticalSymmetry}. We also comment that the $z$-coordinate corresponds to a direction of symmetry generated by the closed conformal Killing vector field.

For the case of $s=0$, the operator~\eqref{LOs} corresponds to a mass ladder operator which maps a solution of a massive Klein-Gordon equation into another solution of that with different mass squared~\cite{Cardoso:2017qmj,Cardoso:2017egd,PhysRevD.106.044052,Katagiri:2021scx,Katagiri:2021xig}. The ladder operator~\eqref{LOs} corresponds to an extension of the mass ladder operator to the case where the massive Klein-Gordon field in ${\rm AdS}_2$ has an additional potential term. In the current case, the important point for the construction is not only that the effective ${\rm AdS}_2$ geometry emerges but also that the potential form inherited from the original black hole geometry belongs to a special class: as is seen in Appendix~\ref{Appendix:ConditionforVerticalSymmetry}, for a scalar field with a generic potential term in ${\rm AdS}_2$, the possible potential form which admits the ladder operator is strongly restricted in the form which is associated with the closed conformal Killing vector field in ${\rm AdS}_2$ (see Eq.~\eqref{generalP}), constraining the form of the corresponding ladder operator at the same time (see Eq.~\eqref{generalD}). In other words, an arbitrary potential term is prohibited by the closed conformal Killing vector field. In this sense, the ladder operator has a geometrical origin. It should then be emphasized that the potential term in Eq.~\eqref{StaticscalarAdS2} admits the ladder operator and is inherited from the property of the original Schwarzschild geometry. This also implies that the linearized Einstein equation around Schwarzschild black holes in the static limit belongs to a special class admitting a ladder operator.

Surprisingly, the expression for the ladder operator~\eqref{laderoperator} with $s=0$ corresponds to that in Eq.~(2.4) in Ref.~\cite{Hui:2021vcv} and that in Eq.~(2.20) in Ref.~\cite{BenAchour:2022uqo} although their constructions are completely different. The expression for the operator~\eqref{laderoperator} with $s=2$  also corresponds to Eqs.~(C.4) and~(C.5) in Ref.~\cite{Hui:2021vcv}, which are heuristically constructed; we derive them and explain their origin in a general context in terms of a detailed analysis based on the geometrical perspective in ${\rm AdS}_2$ in Appendix~\ref{Appendix:ConditionforVerticalSymmetry}.

\subsection{Transformation of static scalar fields by ladder operators }
Now, we act the commutation relation~\eqref{commutationrelation} with $k_-$ in Eq.~\eqref{kpm} on a smooth function~$\phi_\ell(z)$, yielding 
\begin{equation}
\begin{split}
\label{relation}
&\left(\mathcal{D}_{\ell}+\nabla_a\zeta^a\right)\left[\square_{{\rm AdS}_2}-\left(\ell(\ell+1)-\frac{s^2}{z}\right)\right]\phi_\ell=\left[\square_{{\rm AdS}_2}-\left((\ell-1)\ell-\frac{s^2}{z}\right)\right]\mathcal{D}_{\ell}\phi_\ell.
\end{split}
\end{equation}
When the function~$\phi_\ell$ satisfies Eq.~\eqref{StaticscalarAdS2}, the left-hand side of the relation~\eqref{relation} vanishes; thus, obtaining
\begin{equation}
\left[\square_{{\rm AdS}_2}-\left((\ell-1)\ell-\frac{s^2}{z}\right)\right]\mathcal{D}_{\ell}\phi_\ell=0.
\end{equation}
This implies that the function~$\mathcal{D}_{\ell}\phi_\ell$ generated by a transformation~$
\phi_\ell\to \mathcal{D}_{\ell}\phi_\ell
$ is a solution of the equation of motion for $\phi_\ell$ in Eq.~\eqref{StaticscalarAdS2} with $\ell\to\ell-1$. In other words, the action of the ladder operator~$\mathcal{D}_{\ell}$ generates a solution with the multipole index shifted as~$\ell\to\ell-1$. In the same manner, by acting the commutation relation~\eqref{commutationrelation} with $k_+$ in Eq.~\eqref{kpm} on $\phi_\ell$, one can show that the ladder operator generates a function~$\mathcal{D}_{ -\ell-1}\phi_\ell$, which satisfies
\begin{equation}
\left[\square_{{\rm AdS}_2}-\left((\ell+1)(\ell+2)-\frac{s^2}{z}\right)\right]\mathcal{D}_{ -\ell-1}\phi_\ell=0.
\end{equation}
The function~$\mathcal{D}_{-\ell-1}\phi_\ell$ is a solution with the multipole index shifted as $\ell\to \ell+1$. 

Given the commutation relation~\eqref{commutationrelation}, the multiple action of the ladder operators further shifts the index $\ell$ of the solution as follows. When considering $N$-th order multiple actions of the ladder operators on $\phi_\ell$ in Eq.~\eqref{StaticscalarAdS2},
 \begin{equation}
\phi_\ell\to \mathcal{D}_{\ell-N+1}\cdots\mathcal{D}_{\ell-1}\mathcal{D}_{\ell}\phi_\ell,
\end{equation}
we obtain
\begin{equation}
\left[\square_{{\rm AdS}_2}-\left((\ell-N)(\ell-N+1)-\frac{s^2}{z}\right)\right] \mathcal{D}_{\ell-N+1}\cdots\mathcal{D}_{\ell-1}\mathcal{D}_{\ell}\phi_\ell=0.
\end{equation}
This multiple action generates a solution with a shifted index as $\ell\to \ell-N$. In particular, as will be seen later, the $(\ell-s)$-th order action generates the lowest multipole mode from the higher multipole mode indexed with $\ell$. In the same manner, $N$-th order multiple action~$\mathcal{D}_{-\ell-N}\cdots\mathcal{D}_{-\ell-2}\mathcal{D}_{-\ell-1}$ on $\phi_\ell$ generates a solution with the index shifted as $\ell\to\ell+N$.

\subsection{Supersymmetric structure}
The ladder operator relates a static solution with a given~$\ell$ to other static solutions with a shifted multipole index. That means that the transformation generated by the ladder operator keeps the ``shape'' of the equation of motion. We discuss here the underlying structure of the shape invariance, i.e., hidden supersymmetric structure.

Let us consider the system that consists of a pair of fields~$(\phi_\ell,\phi_{\ell\pm1})$ governed by the action,
\begin{equation}
\label{action}
S=\frac{1}{2}\int dz \left[-\left(\Delta\frac{d}{dz}\phi_{\ell}\right)^2-V_{\ell}\phi_{\ell}^2-\left(\Delta\frac{d}{dz}\phi_{\ell\pm1}\right)^2-V_{\ell\pm1}\phi_{\ell\pm1}^2\right],
\end{equation}
where the potentials are given by\footnote{
In terms of $k_\pm$ in Eq.~\eqref{kpm}, the potentials take the forms,
\begin{equation}
\begin{split}
V_\ell=\Delta\left[k_\pm\left(k_\pm+1\right)-\frac{s^2}{z}\right],~~
V_{\ell\pm1}=\Delta\left[\left(k_\pm-1\right)k_\pm-\frac{s^2}{z}\right].
\end{split}
\end{equation}}
\begin{equation}
\begin{split}
\label{VlVlpm1}
V_{\ell}=\Delta\left[\ell\left(\ell+1\right)-\frac{s^2}{z}\right],~~
V_{\ell\pm1}=\Delta\left[\left(\ell\pm1\right)\left(\ell\pm1+1\right)-\frac{s^2}{z}\right].
\end{split}
\end{equation}
The action~\eqref{action} is invariant under the continuous transformation,
\begin{equation}
\label{transformationphiellphiellpm1}
\phi_{\ell}\to \phi_\ell+\epsilon\mathcal{D}_{-k_\pm}\phi_{\ell\pm 1},~~
\phi_{\ell\pm1}\to \phi_{\ell\pm1}+\epsilon \mathcal{D}_{k_\pm}\phi_{\ell},
\end{equation}
where $\epsilon$ is an infinitesimal parameter. We discuss the corresponding conserved quantity in the next subsection. One can 
inductively generalize the above symmetry of the pair~$(\phi_\ell,\phi_{\ell\pm1})$ to that of infinite  pairs~$(\phi_\ell,\phi_{\ell\pm1},\phi_{\ell\pm2},\cdots)$. As will be seen below, the transformation~\eqref{transformationphiellphiellpm1} generates a supersymmetric structure.

We now discuss the detailed property of the above symmetry. By varying the action~\eqref{action} with respect to $\phi_\ell$ and $\phi_{\ell\pm1}$, we obtain the equations of motion for $\phi_\ell$ and $\phi_{\ell\pm1}$, which are consistent with Eq.~\eqref{StaticscalarAdS2},
\begin{equation}
\mathcal{H}_\ell\phi_\ell=0,~~\mathcal{H}_{\ell\pm1}\phi_{\ell\pm1}=0,
\end{equation}
where we have defined the Hamiltonians, 
\begin{equation}
\mathcal{H}_\ell:=-
\Delta\frac{d}{dz}\left(\Delta \frac{d}{dz}
\right)+V_\ell,~~\mathcal{H}_{\ell\pm1}:=-
\Delta\frac{d}{dz}\left(\Delta \frac{d}{dz}\right)+V_{\ell\pm1}.
\end{equation}
Introducing a function,
\begin{equation}
\label{Wkpm}
W_{k_\pm}:=\frac{k_\pm}{2}\left(2z-1-\frac{s^2}{k_\pm^2}\right),
\end{equation}  
the potentials~$V_{\ell}$ and~$V_{\ell\pm1}$ in Eq.~\eqref{VlVlpm1} can be written as
\begin{equation}
\begin{split}
V_{\ell}=W_{k_\pm}^2+\Delta \frac{dW_{k_\pm}}{dz}+\beta_{k_\pm}^2,~~
V_{\ell\pm1}=W_{k_\pm}^2-\Delta \frac{dW_{k_\pm}}{dz}+\beta_{k_\pm}^2,
\end{split}
\end{equation}
where we have defined
\begin{equation}
\label{betakpm}
\beta_{k_\pm}=\frac{k_\pm^2-s^2}{2k_\pm}.
\end{equation}
In terms of $W_{k_\pm}$ in Eq.~\eqref{Wkpm}, the ladder operator takes the form of $\mathcal{D}_{k_\pm}=\Delta\dfrac{d}{dz}-W_{k_\pm}$.
One can then show the following relations:
\begin{equation}
\begin{split}
\label{HDD}
\mathcal{H}_\ell=\mathcal{D}_{-k_\pm}\mathcal{D}_{k_\pm}+\beta_{k_\pm}^2,~~
\mathcal{H}_{\ell\pm1}=\mathcal{D}_{k_\pm}\mathcal{D}_{-k_\pm}+\beta_{k_\pm}^2.
\end{split}
\end{equation}
This implies
\begin{equation}
\begin{split}
\label{intertwiningrelation}
\mathcal{D}_{k_\pm}\mathcal{H}_\ell\phi_\ell&=\mathcal{D}_{k_\pm}\left(\mathcal{D}_{-k_\pm}\mathcal{D}_{k_\pm}+\beta_{k_\pm}^2\right)\phi_\ell=\mathcal{H}_{\ell\pm1}\mathcal{D}_{k_\pm}\phi_\ell=0,\\
\mathcal{D}_{-k_\pm}\mathcal{H}_{\ell\pm1}\phi_{\ell\pm1}&=\mathcal{D}_{-k_\pm}\left(\mathcal{D}_{k_\pm}\mathcal{D}_{-k_\pm}+\beta_{k_\pm}^2\right)\phi_{\ell\pm 1}=\mathcal{H}_{\ell}\mathcal{D}_{-k_\pm}\phi_{\ell\pm1}=0.
\end{split}
\end{equation}
It can be seen that given the solution~$\phi_\ell$ of $\mathcal{H}_\ell$, the function~$\mathcal{D}_{k_\pm}\phi_{\ell}$ is that of $\mathcal{H}_{\ell\pm1}$; given the solution~$\phi_{\ell\pm1}$ of $\mathcal{H}_{\ell\pm1}$, the function~$\mathcal{D}_{-k_\pm}\phi_{\ell\pm1}$ is that of $\mathcal{H}_\ell$.\footnote{This is the emergence of the vertical symmetry in Ref.~\cite{Hui:2021vcv}.} In other words, $\phi_{\ell}$ and $\phi_{\ell\pm1}$ are associated with each other through the transformation generated by the ladder operator. In this sense, $\phi_{\ell}$ and $\phi_{\ell\pm1}$ are a symmetric partner under the transformation by the ladder operators.
Note that for a given~$\phi_\ell$, there exist infinite symmetric partners with different multipoles generated by the multiple action of the ladder operators. This feature corresponds to the supersymmetric structure seen in supersymmetric quantum mechanics~\cite{Cooper:1994eh}.

The above symmetric structure can be understood from the property of the following algebra. Defining matrices,
\begin{equation}
\mathcal{H}=\begin{pmatrix}
\mathcal{H}_\ell-\beta_{k_\pm}^2 & 0 \\
0 & \mathcal{H}_{\ell\pm1}-\beta_{k_\pm}^2\\
\end{pmatrix},~~
\mathcal{Q}=\begin{pmatrix}
0& 0\\
\mathcal{D}_{k_\pm} & 0 \\
\end{pmatrix},~~
\bar{\mathcal{Q}}=\begin{pmatrix}
0& \mathcal{D}_{-k_\pm} \\
0 & 0 \\
\end{pmatrix},
\end{equation}
then, we have
\begin{equation}
\label{verticalsymmetry}
\left[\mathcal{H},\mathcal{Q}\right]=\left[\mathcal{H},\bar{\mathcal{Q}}\right]=O,~~\left\{\mathcal{Q},\bar{\mathcal{Q}}\right\}=\mathcal{H},~~\left\{\mathcal{Q},{\mathcal{Q}}\right\}=\left\{\bar{\mathcal{Q}},\bar{{\mathcal{Q}}}\right\}=O,
\end{equation}
where $O$ is the zero matrix; $[~,~]$ and $\{~,~\}$ denote the commutator and the anti-commutator, respectively. This is known as the supersymmetry algebra, where $\mathcal{D}_{k_\pm}$ and $\mathcal{D}_{-k_\pm}$ correspond to an intertwiner, and $\mathcal{Q}$ and $\bar{\mathcal{Q}}$ correspond to supercharges in supersymmetric quantum mechanics~\cite{Cooper:1994eh}. 

\subsection{Conserved quantity from the supersymmetric structure}
We derive the radially conserved quantity associated with the supersymmetric structure. When performing the transformation~\eqref{transformationphiellphiellpm1}, the invariance of the action~\eqref{action} leads to a conserved quantity,\footnote{There exists another conserved quantity, $\mathcal{\tilde{W}}_{\ell\pm1}=(\mathcal{D}_{k_\pm}\phi_\ell)\left(\Delta\dfrac{d\phi_{\ell\pm1}}{dz}\right)-\left(\Delta\dfrac{d\mathcal{D}_{k_\pm}\phi_\ell}{dz}\right)\phi_{\ell\pm1}$, which is equivalent to the quantity~\eqref{Well} for the fields satisfying the equations of motion.
}
\begin{equation}
\begin{split}
\label{Well}
\mathcal{W}_\ell:=\left(\mathcal{D}_{-k_\pm}\phi_{\ell\pm1}\right)\left(\Delta\frac{d\phi_{\ell}}{dz}\right)-\left(\Delta\frac{d\mathcal{D}_{-k_\pm}\phi_{\ell\pm1}}{dz}\right)\phi_{\ell}.
\end{split}
\end{equation}
One can show the radial conservation of $\mathcal{W}_\ell$ for the fields that satisfy the equations of motion,~$\mathcal{H}_\ell \phi_\ell=0$ and $\mathcal{H}_{\ell\pm1}\phi_{\ell\pm1}=0$. In other words, the quantity~${\cal W}_\ell$ is conserved in the direction generated by the closed conformal Killing vector field.

We discuss the property of the conserved quantity in the radial direction. Let us consider the pair of the lowest multipole mode~$\phi_s$ for a given spin weight~$s$ and an originally ``unphysical'' mode~$\phi_{s-1}$. From the explicit forms of the ladder operator,~$\mathcal{D}_{\mp s}=z(z-1)\dfrac{d}{dz}\pm s(z-1)$, it follows that  $\mathcal{D}_{-s}\phi_{s-1}|_{z\to1}=0$ and $\mathcal{D}_s\phi_s|_{z\to1}=0$ for $\phi_{s-1}$ and $\phi_s$ with the regularity condition at $z=1$, while this is not the case for the logarithmically divergent solutions. This is a unique property for the pair with $\phi_{s-1}$, and is not the case for pairs without it, e.g., $(\phi_s,\phi_{s+1})$. The radial conservation of $\mathcal{W}_s$ then leads to
\begin{equation}
\left(\mathcal{D}_{-s}\phi_{s-1}\right)
\frac{d}{dz}\left[z^s\mathcal{D}_s\phi_s\right]=0.
\end{equation}
This means that with the general solution of $\phi_{s-1}$ prepared, 
the conservation of $\mathcal{W}_s$ is equivalent to that of the following quantity:
\begin{equation}
\label{Y2}
Y_s:=z^s\mathcal{D}_s\phi_s.
\end{equation}
From the observations above, the radially conserved quantity~$Y_s$ identically vanishes for $\phi_s$ with regularity at $z=1$, while that is nonzero for the logarithmically divergent solution. This shows that $\mathcal{D}_s\phi_s$ itself identically vanishes for $\phi_s$ with regularity. The quantity~$Y_s$ is proportional to the induced multipole moment~\cite{Damour:2009vw,Binnington:2009bb} as will be seen in Sec.~\ref{Sec:VanishingLovefromSymmetry}. The vanishing of $D_{s}\phi_s$ under the regularity condition is similar to the phenomenon known as ``unbroken supersymmetry'' in the context of supersymmetric quantum mechanics~\cite{Cooper:1994eh}.\footnote{One can show from another conserved quantity~$\mathcal{\tilde{W}}_{s-1}$ that $\mathcal{D}_{-s}\phi_{s-1}$ also identically vanishes; this is also similar to unbroken supersymmetry~\cite{Cooper:1994eh}.} 

One can obtain a radially conserved quantity for general multipoles. The ladder operators map a given $\phi_\ell$ into the lowest multipole mode, $\tilde{\phi}_s:=\mathcal{D}_{s+1}\cdots\mathcal{D}_{\ell-1}\mathcal{D}_\ell\phi_\ell$, thereby yielding a conserved quantity in the radial direction,
\begin{equation}
\label{Yell}
Y_\ell:=z^s\mathcal{D}_s\tilde{\phi_s}.
\end{equation}
The quantity~$Y_\ell$ is proportional to the induced multipole moment~\cite{Damour:2009vw,Binnington:2009bb}.

\section{No static response of Schwarzschild black hole from hidden symmetry}
\label{Sec:VanishingLovefromSymmetry}
In this section, we discuss no static response of the Schwarzschild black hole to the spin-$s$ field perturbation in terms of the supersymmetric structure. 
In Appendix~\ref{Appendix:tidalLovenumberofPolar}, we show that for the case of the polar-type tidal field perturbation.

\subsection{Lowest multipole mode case}
First, we show the vanishing of the Love and dissipation numbers for the lowest multipole mode~$\phi_{s}$ with a given spin weight~$s$, by exploiting the radial conservation of $Y_\ell$ in Eq.~\eqref{Y2}. For the general solution of $\phi_s$, we have
\begin{equation}
\begin{split}
    \label{ConservedY2}
{Y}_{s}=&z^s\mathcal{D}_s\phi_s\\
=&z^{2s}\Delta\frac{d}{dz}\left(z^{-s}\phi_s\right).
\end{split}
\end{equation}
The radial conservation of ${Y}_s$ includes horizontal symmetry found by direct inspection of the Regge-Wheeler equation in Ref.~\cite{Hui:2021vcv}.

Without knowledge of the exact solutions of the perturbations, we instead use the asymptotic solutions at the horizon and large distances, i.e.,
\begin{equation}
\left.\phi_s\right|_{z\to 1}\sim {\rm const.},~~\ln\left(1-1/z\right),
\end{equation}
and
\begin{equation}
\left.\phi_s\right|_{z\gg 1}\sim z^s,~~z^{-s-1}.
\end{equation}
The translational conservation of $Y_s$ of Eq.~\eqref{ConservedY2} tells us the asymptotic behavior of the solution at large distances under the requirement of the regularity at the horizon as follows. The evaluation of $Y_s$ in Eq.~\eqref{ConservedY2} for $\phi_s|_{z\to 1}\sim {\rm const}.$ yields $Y_s=0$ at the horizon~$(\Delta=0)$, as we have already known, for $\phi_s|_{z\gg 1}\sim z^{s}$ it also leads to $Y_s=0$. 
On the other hand, the values of $Y_s$ are nonzero for $\phi_s|_{z\to 1}\sim \ln(1-1/z)$ and $\phi_s|_{z\gg 1}\sim z^{-s-1}$; then, $Y_s$ is included in the factor in front of the decaying solution and is proportional to the induced multipole moment~\cite{Damour:2009vw,Binnington:2009bb}. Therefore, the solution regular~(logarithmically divergent) at the horizon must connect to a purely growing~(decaying, respectively) solution of $z$ at large distances. This shows that for $\phi_s$, the spin-$s$-field Love and dissipation numbers of the Schwarzschild black hole vanish.

\subsection{General multipole case}
Next, we extend the discussion above to the general multipole case by using the ladder operators. As stated at the end of Sec.~\ref{Sec:HiddenSymmetryfromAdS2}, the ladder operators map a given mode~$\phi_\ell$ into the lowest multipole mode~$\tilde{\phi}_s=\mathcal{D}_{s+1}\cdots\mathcal{D}_{\ell-1}\mathcal{D}_\ell\phi_\ell$, giving rise to a conserved quantity in the radial direction,
\begin{equation}
\begin{split}
\label{ConservedYell}
{Y}_{\ell}=&z^s\mathcal{D}_s\tilde{\phi}_s\\
=&z^{2s}\Delta\frac{d}{dz}\left(z^{-s}\tilde{\phi}_s\right),
\end{split}
\end{equation}
which is the same form as $Y_s$ in Eq.~\eqref{ConservedY2}.

Asymptotic solutions of $\phi_\ell$ at the horizon and large distances are, respectively,
\begin{equation}
\left.\phi_\ell\right|_{z\to1}\sim {\rm const.},~~\ln\left(1-1/z\right),
\end{equation}
and
\begin{equation}
\label{asymptoticbehavior:Schwarzschild}
\left.\phi_\ell\right|_{z\gg1}\sim z^{\ell},~~z^{-\ell-1}.
\end{equation}
Noting that the ladder operators have no logarithmic term, the solution regular at the horizon,~$\phi_\ell|_{z\to1}\sim {\rm const}.$, should give rise to $\tilde{\phi}_s|_{z\to1}\sim {\rm const}.$, showing the vanishing of the conserved quantity~$Y_\ell$. On the other hand, the solution logarithmically divergent at the horizon, i.e., $\phi_\ell|_{z\to 1}\sim \ln(1-1/z)$, leads to $\tilde{\phi}_s|_{z\to 1}\sim \ln(1-1/z)$, giving the nonzero~$Y_\ell$.

From the asymptotic form of the ladder operators at large distances,
\begin{equation}
\label{AsymptoticD}
\mathcal{D}_{\ell}=
z^2\left(1+\mathcal{O}\left(1/z\right)\right)\dfrac{d}{dz}-\ell z\left(1+\mathcal{O}\left(1/z\right)\right)~~{\rm for}~~z\gg 1,
\end{equation}
it follows that the ladder operator~$\mathcal{D}_\ell$ cancels out the leading term of $\phi_\ell|_{z\gg1}\sim z^{\ell}$. Hence, the leading behavior of $\mathcal{D}_\ell \phi_\ell|_{z\gg1}$ comes from the subleading term of $\phi_\ell|_{z\gg1}$, leading to $\mathcal{D}_\ell \phi_\ell|_{z\gg1}\sim z^{\ell-1}$. In the same manner, one can show $\mathcal{D}_{\ell-1}\mathcal{D}_{\ell} \phi_\ell|_{z\gg1}\sim z^{\ell-2}$. This indicates that $(\ell-s)$-th order action of the ladder operators on $\phi_\ell|_{z\gg1}\sim z^{\ell}$ leads to $\tilde{\phi}_s|_{z\gg1}\sim z^s$, giving the vanishing of $Y_\ell$. On the other hand, the ladder operator on $\phi_\ell|_{z\gg1}\sim z^{-\ell-1}$ leads to $\mathcal{D}_\ell\phi_\ell|_{z\gg1}\sim z^{-\ell}$. Hence, the $(\ell-s)$-th order action leads to $\tilde{\phi}_s|_{z\gg1}\sim z^{-s-1}$, giving nonzero~$Y_\ell$,  which is proportional to the induced multipole moment~\cite{Damour:2009vw,Binnington:2009bb}.

To summarize the above analysis, the solution regular at the horizon, i.e., $\phi_\ell|_{z\to1}\sim {\rm const}.$, and the solution purely growing at large distances, i.e., $\phi_\ell|_{z\gg 1}\sim z^{\ell}$, give the vanishing~$Y_\ell$, while the solution divergent at the horizon, i.e., $\phi_\ell|_{z\to1}\sim \ln(1-1/z)$, and the solution purely decaying at large distances, i.e., $\phi_\ell|_{z\gg 1}\sim z^{-\ell-1}$, lead to the nonzero~$Y_\ell$. This shows that the solution regular at the horizon connects to that purely growing at the distant region, and does not contain the decaying term, i.e., the vanishing of spin-$s$-field Love and dissipation numbers for all~$\ell$.

\section{Vanishing Love of Kerr black holes from hidden symmetry}
\label{Sec:VanishingLoveofKerrBH}
We here study the vanishing of Love numbers of the Kerr black hole for static spin-$s$ field perturbations in terms of a hidden supersymmetric structure. We first reduce the static perturbation into a set of infinite static scalar fields in ${\rm AdS}_2$. It is then found that there exists the hidden supersymmetric structure from spacetime conformal symmetry of the reduced geometry in the parallel manner as in the Schwarzschild black hole case. We show no static response of the Kerr black hole in terms of the associated conserved quantity. We also discuss the vanishing Love numbers of the Kerr black hole with the nonzero dissipation numbers for the non-axisymmetric perturbations.

\subsection{Spin-$s$-field Love numbers of Kerr black holes}
Linear perturbations to the Kerr black hole are governed by the Teukolsky equation~\cite{PhysRevLett.29.1114}. The radial Teukolsky equation for the spin-$s$ field perturbation in the static limit is given by
\begin{equation}
\label{staticTeukolskyeq}
\Delta^{-s}\frac{d}{dz}\left(\Delta^{s+1}\frac{d}{dz}\Phi_{\ell m}^{(s)}(z)\right)+\left(\frac{m^2\chi^2+i m\chi s(2z-1)}{\Delta}-\ell(\ell+1)+s(s+1)\right)\Phi_{\ell m}^{(s)}(z)=0,
\end{equation}
where $\ell$ and $m$ are integers such that $\ell\ge s$ and $|m|\le \ell$. The dimensionless parameter~$\chi\in[0,1)$ denotes a spin of the black hole and  
\begin{equation}
\label{Delta:Kerr}
\Delta=z(z-1).
\end{equation}
The dimensionless radial coordinate~$z\in (1,\infty)$ is related to the areal coordinate~$r$ as
\begin{equation}
\label{z:Kerr}
z=\frac{r-r_-}{r_+-r_-},
\end{equation}
where $r_+$ and $r_-$ are the radii of the event and Cauchy horizons, respectively. In this coordinate, the event horizon and infinity are located at $z=1$ and $z=\infty$, respectively.

The static response of the Kerr black hole is characterized by the asymptotic behavior of the field at large distances under the requirement of the smoothness of a function~$\Delta^2\Phi_{\ell m}^{(s)}$ at the horizon~$z=1$~\cite{LeTiec:2020spy,LeTiec:2020bos}:\footnote{The requirement above corresponds to eliminating of a logarithmic contribution from the asymptotic solution near the horizon. The presence of the logarithmic contribution gives blowups of the second-order derivative of $\Delta^2\Phi_{\ell m}^{(s)}$ at the horizon~\cite{LeTiec:2020spy,LeTiec:2020bos}. The smoothness condition also corresponds to no outgoing-wave boundary condition for time-dependent fields in the zero-frequency limit~\cite{LeTiec:2020spy,LeTiec:2020bos}.}
\begin{equation}
\left.\Phi_{\ell m}^{(s)}\right|_{z\gg1}\propto z^{\ell-s}\left[1+\mathcal{O}(1/z)+\kappa_{\ell m}^{(s)} z^{-2\ell-1}\left(1+\mathcal{O}(1/z)\right)\right].
\end{equation}
Note that in general the coefficient~$\kappa_{\ell m}^{(s)}$ is complex based on the Newtonian analogy for rotating bodies. The real part of $\kappa_{\ell m}^{(s)}$ is called the (conservative) spin-$s$-field Love numbers. The dissipation numbers can be read off from the imaginary part~\cite{Chia:2020yla,Charalambous:2021mea}. 
For spin-$s$ field perturbations, the Kerr black hole has zero conservative Love numbers but has nonzero dissipation numbers in general~\cite{Chia:2020yla,Charalambous:2021mea}. In the case of the axisymmetric static perturbation~$m=0$, the dissipation numbers also vanish~\cite{LeTiec:2020spy,LeTiec:2020bos}.

\subsection{Reduction to ${\rm AdS}_2$}
We first reduce the static spin-$s$ field perturbation into a two-dimensional problem. Introducing a new variable,
\begin{equation}
\phi_{\ell m}^{(s)}(z):=\Delta^{s/2}\Phi_{\ell m}^{(s)},
\end{equation}
the static Teukolsky equation~\eqref{staticTeukolskyeq} can be rewritten as
\begin{equation}
\frac{d}{dz}\left(\Delta\frac{d}{dz}\phi_{\ell m}^{(s)}\right)-\left(\ell(\ell+1)-\frac{4m^2\chi^2-s^2}{4\Delta}+im\chi s \frac{1-2z}{\Delta}\right)\phi_{\ell m}^{(s)}=0.
\end{equation}
This equation can be identified as an equation of motion for a static scalar field in ${\rm AdS}_2$, whose line element is given by 
Eq.~\eqref{AdS2metric}. We thus reduce the static perturbation to the Kerr black hole into a set of infinite static scalar fields in ${\rm AdS}_2$, which satisfy
\begin{equation}
\label{staticfieldinAdS2:Kerr}
\left[\Box_{{\rm AdS}_2}-\left(\ell(\ell+1)-\frac{4m^2\chi^2-s^2}{4\Delta}+im\chi s \frac{1-2z}{\Delta}\right)\right]\phi_{\ell m}^{(s)}=0,
\end{equation}
where the d'Alembertian~$\Box_{{\rm AdS}_2}$ is given by Eq.~\eqref{BoxAdS2}. Note that even in the non-rotating limit~$\chi \to0$, Eq.~\eqref{staticfieldinAdS2:Kerr} does not reproduce Eq.~\eqref{StaticscalarAdS2} because the Teukolsky equation is not smoothly reduced to the Regge-Wheeler equation.

\subsection{Supersymmetric structure and radially conserved quantities}
We here show the existence of a supersymmetric structure and derive the associated conserved quantity in the radial direction. As shown in Appendix~\ref{Appendix:ConditionforVerticalSymmetry}, in the current system, we have a ladder operator arising from spacetime conformal symmetry of ${\rm AdS}_2$,
\begin{equation}
\label{LOs:Kerr}
\mathcal{D}_{k_\pm}=\Delta\frac{d}{dz}-\frac{k_\pm}{2}\left(2z-1-i\frac{2 m\chi s}{k_\pm^2}\right),~~k_+=-\ell-1,~~k_-=\ell,
\end{equation}
which shifts $\ell$ into $\ell\pm1$ in Eq.~\eqref{staticfieldinAdS2:Kerr}. This arises from the Kerr geometry for the same reason as the ladder operator~\eqref{LOs} for the Schwarzschild black hole does (see also Appendix~\ref{Appendix:ConditionforVerticalSymmetry}). Note that the construction of the ladder operator is coordinate-independent. The ladder operator~\eqref{LOs:Kerr} corresponds to the operators in Eqs.~(3.7) and~(D.2) in Ref.~\cite{Hui:2021vcv}, and the operator with $\omega=0$ in Eqs.~(19a) and~(19b) in Ref.~\cite{Hui:2022vbh}.

To see that the operator~\eqref{LOs:Kerr} is a generator of a hidden supersymmetric structure, let us consider a pair of the fields,~$(\phi_{\ell m}^{(s)},\phi_{\ell\pm1 m}^{(s)})$, governed by the action,
\begin{equation}
S=\frac{1}{2}\int dz\left[-\left(\Delta\frac{d}{dz}\phi_{\ell m }\right)^2-V_{\ell m}\phi_{\ell  m}^2-\left(\Delta\frac{d}{dz}\phi_{\ell\pm1m }\right)^2-V_{\ell\pm1 m}\phi_{\ell\pm1 m}^2\right],
\end{equation}
where
\begin{equation}
\begin{split}
V_{\ell m}=&\Delta\left[\ell(\ell+1)-\frac{4m^2\chi^2-s^2}{4\Delta}+im\chi s \frac{1-2z}{\Delta}\right],\\
V_{\ell\pm1 m}=&\Delta\left[\left(\ell\pm1\right)(\ell\pm1+1)-\frac{4m^2\chi^2-s^2}{4\Delta}+im\chi s \frac{1-2z}{\Delta}\right].
\end{split}
\end{equation}
Henceforth, we omit the superscript~$s$ of the variables. The equations of motion for $\phi_{\ell m}$ and $\phi_{\ell\pm1m}$ are given by
\begin{equation}
\mathcal{H}_{\ell m}\phi_{\ell m }=0,~~\mathcal{H}_{\ell\pm1 m}\phi_{\ell\pm1 m}=0,
\end{equation}
where the Hamiltonians are defined as
\begin{equation}
\mathcal{H}_{\ell m}:=-\Delta\frac{d}{dz}\left(\Delta\frac{d}{dz}\right)+V_{\ell m},~~\mathcal{H}_{\ell\pm1 m}:=-\Delta\frac{d}{dz}\left(\Delta\frac{d}{dz}\right)+V_{\ell\pm1 m},
\end{equation}
which are consistent with Eq.~\eqref{staticfieldinAdS2:Kerr}. One can then show
\begin{equation}
\mathcal{H}_{\ell m }=\mathcal{D}_{-k_\pm}\mathcal{D}_{k_\pm}+\beta_{-k_\pm}\beta_{k_\pm},~~\mathcal{H}_{\ell\pm1 m }=\mathcal{D}_{k_\pm}\mathcal{D}_{-k_\pm}+\beta_{k_\pm}\beta_{-k_\pm},
\end{equation}
with
\begin{equation}
\beta_{k_\pm}=\left(\frac{1}{2}+i\frac{m\chi}{k_\pm}\right)\left(s+k_\pm\right).
\end{equation}
Here, we have
\begin{equation}
\begin{split}
\mathcal{D}_{k_\pm}\mathcal{H}_{\ell m}\phi_{\ell m }&=\mathcal{H}_{\ell\pm1 m}\mathcal{D}_{k_\pm}\phi_{\ell m }=0,\\
\mathcal{D}_{-k_\pm}\mathcal{H}_{\ell\pm1 m}\phi_{\ell\pm1m }&=\mathcal{H}_{\ell m}\mathcal{D}_{-k_\pm}\phi_{\ell\pm1m }=0,
\end{split}
\end{equation}
which has the same structure as Eq.~\eqref{intertwiningrelation} does. Therefore, $\phi_{\ell m}$ and $\phi_{\ell\pm1m }$ are a symmetric partner under the transformation generated by the ladder operator, which can be understood from the property of the supersymmetry algebra. Note that the supersymmetric structure can be generalized to that of infinite pairs~$(\phi_{\ell m },\phi_{\ell\pm1m }, \phi_{\ell\pm2 m},\cdots)$.

In the same manner as for the Schwarzschild black hole (Sec.~\ref{Sec:HiddenSymmetryfromAdS2}), the pair of the lowest multipole and an originally ``unphysical'' mode, i.e., $(\phi_{s m}, \phi_{s-1 m})$, gives rise to a constant in the radial direction,
\begin{equation}
\label{Ps:Kerr}
P_{s m}:=\left(\frac{z-1}{z}\right)^{-i  m\chi}\Delta^{s/2}\mathcal{D}_{s}\phi_{s m}.
\end{equation}
For general multipoles, one can also obtain radial constants by mapping $\phi_{\ell m}$ into a lowest multipole~$\tilde{\phi }_{sm}:=\mathcal{D}_{s+1}\cdots\mathcal{D}_{\ell-1}\mathcal{D}_{\ell}\phi_{\ell m}$:
\begin{equation}
\label{Pell:Kerr}
P_{\ell m}:=\left(\frac{z-1}{z}\right)^{-i  m\chi}\Delta^{s/2}\mathcal{D}_s\tilde{\phi}_{s m}.
\end{equation}
The quantity~$P_{\ell m}$ can be interpreted as the radially conserved quantity associated with the supersymmetric structure of the static scalar field in ${\rm AdS}_2$. 

For the case of~$m\neq 0$, $\phi_{\ell m}$ of $|m|>\ell$ can be generated by the ladder operator. In terms of the original perturbation field to the Kerr black hole, the combination of the radial function of $|m|>\ell$ with the corresponding angular function is singular with respect to the angular variables. This means that the structure generated by the ladder operator~\eqref{LOs:Kerr} with~$m\neq 0$ and the constant~$P_{\ell m}$ are not prescribed under the regularity condition to the angle for the original perturbation field. Although the structure is nothing else but a reflection of the mathematical property of the current four-dimensional system, the interpretation is controversial~(see discussion in Refs.~\cite{Hui:2021vcv, Hui:2022vbh,Charalambous:2022rre}). On the other hand, for~$m=0$, no such solutions are generated. Therefore, the structure generated by the ladder operator~\eqref{LOs:Kerr} with $m=0$ and the radially conserved quantity~$P_{\ell 0}$ are prescribed under the regularity condition to the angle for the original perturbation field.

In the following, we show no static response, i.e., vanishing of both Love and dissipation numbers, of the Kerr black hole for the axisymmetric field~$(m=0)$ with the radially conserved quantity~$P_{\ell 0}$. We also discuss the vanishing Love numbers  of the Kerr black hole with the nonzero dissipation numbers for the non-axisymmetric fields~$(m\neq0)$ in terms of the constant in the radial direction, $P_{\ell m}$, even though that is not prescribed under the regularity condition to the angle. We leave a further discussion on the geometrical interpretation of the aforementioned structure of the non-axisymmetric perturbation field in outlook.

\subsection{No static response for axisymmetric fields~$(m=0)$}
We show the vanishing of the Love and dissipation numbers of the Kerr black holes for the axisymmetric mode~$(m=0)$ from the radially conserved quantities $P_{s m=0}$ in Eq.~\eqref{Ps:Kerr} and $P_{\ell m=0}$ in Eq.~\eqref{Pell:Kerr}. For a given spin weight~$s$, we have two linearly independent asymptotic solutions of Eq.~\eqref{staticfieldinAdS2:Kerr} at $z=1$ and $z\gg1$, respectively:
\begin{equation}
\begin{split}
\left.\phi_{\ell0}^{(s=0)}\right|_{z\to1}\sim & {\rm const}.,~\ln(z-1),~~\left.\phi_{\ell0}^{(s=1,2)}\right|_{z\to1}\sim (z-1)^{s/2},~(z-1)^{-s/2},
\end{split}
\end{equation}
and
\begin{equation}
\label{asymptoticbehavior:Kerr}
\left.\phi_{\ell0}\right|_{z\gg1}\sim z^\ell,~z^{-\ell-1}.
\end{equation}
The Love and dissipation numbers can be read off from the asymptotic behaviors at large distances under the requirement of the smoothness of $\Delta^2\Phi_{\ell0}$ at~$z=1$~\cite{LeTiec:2020spy,LeTiec:2020bos}:
\begin{equation}
\begin{split}
\label{reg:Kerr}
\left.\phi_{\ell0}^{(s=0)}\right|_{z\to1}\sim  {\rm const}.,~~\left.\phi_{\ell0}^{(s=1,2)}\right|_{z\to1}\sim& (z-1)^{s/2}.
\end{split}
\end{equation}
The other asymptotic solution, i.e., $ \phi_{\ell0}^{(s=1,2)}|_{z\to1}\sim (z-1)^{-s/2}$, has a logarithmic contribution in the subleading term, leading to the blowups of the second-order derivative of $\Delta^2\Phi_{\ell0}^{(s=1,2)}$ at the horizon.

We show the vanishing of the Love and dissipation numbers by using the radially conserved quantities $P_{s m=0}$ and $P_{\ell m=0}$ of Eqs.~\eqref{Ps:Kerr} and \eqref{Pell:Kerr}, respectively, in the same manner as in the Schwarzschild black hole case in Sec.~\ref{Sec:VanishingLovefromSymmetry}: for the axisymmetric scalar field~($m=0,s=0$), the system~\eqref{staticfieldinAdS2:Kerr} and the ladder operator~\eqref{LOs:Kerr} are the same forms as those for the Schwarzschild black hole, i.e., Eqs.~\eqref{StaticscalarAdS2} and~\eqref{laderoperator} with $s=0$, respectively, implying that $\phi_{\ell0}^{(0)}$ satisfying the condition~\eqref{reg:Kerr} is purely growing at large distances~$z\gg1$. Thus, we immediately conclude that the scalar-field Love and dissipation numbers vanish for all the axisymmetric modes~$\phi_{\ell 0}^{(0)}$, i.e., $\kappa_{\ell0}^{(0)}=0$. 

For the vector~$(s=1)$ and tidal fields~$(s=2)$, the asymptotic behaviors of $\phi_{\ell0}$ in Eq.~\eqref{asymptoticbehavior:Kerr} and the ladder operator~\eqref{LOs:Kerr} at large distances~$z\gg1$ are the same forms as those for the Schwarzschild black hole, i.e., Eqs.~\eqref{asymptoticbehavior:Schwarzschild} and~\eqref{AsymptoticD}, implying that the radially conserved quantity~$P_{\ell m=0}$ vanishes for the purely growing solution~$\phi_{\ell0}|_{z\gg1}\sim z^\ell$, while does not for the purely decaying solution~$\phi_{\ell0}|_{z\gg1}\sim z^{-\ell-1}$. Furthermore, for the lowest multipole mode, the quantity $P_{s m=0}$ of Eq.~\eqref{Ps:Kerr} vanishes for $\phi_{s0}$ satisfying the condition~\eqref{reg:Kerr} at $z=1$, while $P_{s 0}$ can have a nonzero value for the other asymptotic solution at $z=1$. These imply that $\phi_{s0}$ satisfying the condition~\eqref{reg:Kerr} is purely growing at large distances; therefore, the Love and dissipation numbers for $\phi_{s0}$ vanish, i.e., $\kappa_{s0}=0$. For general multipoles, one can also show that the radially conserved quantity $P_{\ell m=0}$ of Eq.~\eqref{Pell:Kerr} vanishes for $\phi_{\ell0}$ with the condition~\eqref{reg:Kerr} at $z=1$, while does not for the other asymptotic solution at $z=1$, implying the vanishing of the spin-$s$-field Love and dissipation numbers for all the axisymmetric modes~$\phi_{\ell0}$, i.e., $\kappa_{\ell 0}^{(s)}=0$. 

\subsection{Discussion: non-axisymmetric field case~$(m\neq0)$}
We discuss the vanishing of the Love numbers for the non-axisymmetric modes~$(m\neq 0)$ in terms of the radial constants, $P_{s m}$ and $P_{\ell m}$~in Eqs.~\eqref{Ps:Kerr} and~\eqref{Pell:Kerr}. Although the structure generated by the ladder operator and the constant~$P_{\ell m}$ are not prescribed under the regularity condition to the angle in terms of the perturbation field to the Kerr background, we show the vanishing of the Love numbers in the following.

We introduce a new variable,
\begin{equation}
\label{newF}
F_{\ell m}(z):=\Delta^{s/2}\left(\frac{z}{z-1}\right)^{i m \chi} \phi_{\ell m}.
\end{equation}
The asymptotic behaviors of $F_{\ell m}$ at $z=1$ and $z\gg1$ are, respectively, given by
\begin{equation}
\label{Fellmwithnonaxisymmetric}
\left.F_{\ell m}\right|_{z\to1}\sim{\rm const}.,~(z-1)^{-2im\chi+s},
\end{equation}
and
\begin{equation}
\label{Fatlargez}
\left.F_{\ell m}\right|_{z\gg1}\sim z^{\ell+s},~z^{-\ell+s-1}.
\end{equation}
The requirement of the smoothness of $\Delta^2\Phi_{\ell m}$ at $z=1$ as in Refs.~\cite{LeTiec:2020spy,LeTiec:2020bos} corresponds to
\begin{equation}
\label{phiellmathorizon}
\left.F_{\ell m}\right|_{z\to1}\sim {\rm const}.,
\end{equation}
whose asymptotic behavior at large distances~$z\gg1$ determines the Love and dissipation numbers~\cite{Chia:2020yla,Charalambous:2021mea,Hui:2021vcv,Hui:2022vbh}.

We first discuss the lowest multipole case~$\phi_{sm}$. The constant in the radial direction, $P_{s m}$ of Eq.~\eqref{Ps:Kerr}, can be rewritten in terms of $F_{s m}$ as
\begin{equation}
\label{Psm:Kerr}
P_{sm}=z\left(z-1\right)\frac{d}{dz}F_{s m}+\left[2im \chi-s(2z-1)\right]F_{s m}.
\end{equation}
It follows from the asymptotic behaviors of $F_{sm}$ given in Eq.~\eqref{Fellmwithnonaxisymmetric} that $P_{sm}$ is nonzero for $F_{sm}$ satisfying the condition~\eqref{phiellmathorizon} and vanishes identically for the other solution in Eq.~\eqref{Fellmwithnonaxisymmetric} from the constant in the radial direction. Furthermore, both the asymptotic solutions at large distances~$z\gg1$, i.e., those of Eq.~\eqref{Fatlargez}, lead to nonzero~$P_{sm}$, implying that $F_{sm}$ satisfying the condition~\eqref{phiellmathorizon} behaves as
\begin{equation}
\label{asymptoticphism:Kerr}
\left.F_{sm}\right|_{z\gg1}\propto z^{2s}\left[1+\mathcal{O}(1/z)+\kappa_{s m}z^{-2s-1}\left(1+\mathcal{O}(1/z)\right)\right],
\end{equation}
for a nonzero coefficient~$\kappa_{sm}$. Note that $P_{sm}$ corresponds to an overall factor of $F_{sm}$ satisfying the condition~\eqref{phiellmathorizon} and is proportional to the tidal moment~\cite{Damour:2009vw,Binnington:2009bb}.

The vanishing of the spin-$s$-field Love number can be shown as follows. The nonzero~$P_{sm}$ implies that $F_{sm}$ is a $2s$-th order finite polynomial of $z-1$, i.e., $F_{sm}=\sum_{j=0}^{2s}\alpha_{j}(z-1)^{j}$, which satisfies the boundary condition~\eqref{phiellmathorizon}.\footnote{Substituting~$F_{sm}=\sum_{j=0}^\infty \alpha_j (z-1)^j$ into Eq.~\eqref{Psm:Kerr}, one obtains a recurrence relation~$(j+1-s+2im\chi)\alpha_{j+1}=-(j-2s)\alpha_{j}$ with~$\alpha_0=P_{sm}/(2im\chi-s)$, which implies $\alpha_{j\le2s}\neq0$ and $\alpha_{j\ge2s+1}=0$ if $P_{sm}\neq0$.} With the explicit forms of~$\alpha_{j}$, one can show that the polynomial is proportional to a hypergeometric series that defines the Gaussian hypergeometric function~\cite{abramowitz+stegun}:
\begin{equation}
\label{FandF}
F_{sm}
\propto~_2F_1\left(-2s, 1;1-s+2im\chi;1-z\right),
\end{equation}
where $_2F_1(~,~;~;1-z)$ is the Gaussian hypergeometric function around $z=1$. Performing the transformation for $_2F_1$~\cite{abramowitz+stegun},
\begin{equation}
\begin{split}
\label{transformaionfor2F1}
&~_2F_1(\alpha,\beta;\gamma;1-z)\\
&=\frac{\Gamma(\gamma)\Gamma(\beta-\alpha)}{\Gamma(\beta)\Gamma(\gamma-\alpha)}z^{-\alpha}\left[~_2F_1\left(\alpha,\gamma-\beta;\alpha-\beta+1;\frac{1}{z}\right)+\kappa z^{\alpha-\beta}~_2F_1\left(\beta,\gamma-\alpha;\beta-\alpha+1;\frac{1}{z}\right)\right],
\end{split}
\end{equation}
with 
\begin{equation}
\label{kappasm}
\kappa=\frac{\Gamma(\beta)\Gamma(\alpha-\beta)\Gamma(\gamma-\alpha)}{\Gamma(\alpha)\Gamma(\gamma-\beta)\Gamma(\beta-\alpha)},
\end{equation}
and
\begin{equation}
\alpha=-2s,~~\beta=1,~~\gamma=1-s+2im\chi,
\end{equation}
one can see that the asymptotic behavior of $F_{sm}$ at large distances takes the form of Eq.~\eqref{asymptoticphism:Kerr} because $_2F_1(~,~;~;1/z)|_{z\gg1}=1+\mathcal{O}(1/z)$. In particular, we further rewrite Eq.~\eqref{kappasm} to~\cite{abramowitz+stegun}
\begin{equation}
\label{ksm}
\kappa=\kappa_{sm}:=-im\chi\frac{\left(-1\right)^{s}}{2\left(2s+1\right)!}\prod_{j=1}^s\left[j^2\left(1-\chi^2\right)+m^2\chi^2\right],
\end{equation}
which gives an analytic expression for the dissipation number~\cite{Chia:2020yla,Charalambous:2021mea,Hui:2022vbh}, explicitly showing the vanishing of the spin-$s$-field Love number for the lowest multipole, i.e., ${\rm Re}[\kappa_{sm}^{(s)}]=0$.

We next discuss the higher multipole case. The ladder operator~$\mathcal{D}_{k_-}$ in Eq.~\eqref{LOs:Kerr} keeps the asymptotic behavior of $\phi_{\ell m}$ at $z=1$ unchanged. This implies that $\phi_{\ell m}$ compatible with the condition~\eqref{phiellmathorizon} can have nonzero $P_{\ell m}$ of Eq.~\eqref{Pell:Kerr}, while the other solution has $P_{\ell m}=0$. In addition, the fact that the ladder operator~$\mathcal{D}_{k_-}$ maps the asymptotic solutions of $\phi_{\ell m}$ compatible with Eq.~\eqref{Fatlargez} at $z\gg1$ into those with $\ell\to\ell-1$ implies that $P_{\ell m}$ is nonzero for them. Therefore, $F_{\ell m}$ with the condition~\eqref{phiellmathorizon} behaves as
\begin{equation}
\label{asymptoticphiellm:Kerr}
\left.F_{\ell m}\right|_{z\gg1}\propto z^{\ell+s}\left[1+\mathcal{O}(1/z)+\kappa_{\ell m}z^{-2\ell-1}\left(1+\mathcal{O}(1/z)\right)\right],
\end{equation}
for a nonzero coefficient~$\kappa_{\ell m}$. Note that $P_{\ell m}$ for $F_{\ell m}$ satisfying the condition~\eqref{phiellmathorizon} is an overall factor and is proportional to the tidal moment~\cite{Damour:2009vw,Binnington:2009bb}.

The constant~$P_{\ell m}$~\eqref{Pell:Kerr} implies that $F_{\ell m}$ can be expressed as an $(\ell+s)$-th order finite polynomial of $z-1$. The explicit form is proportional to a hypergeometric series that defines the Gaussian hypergeometric function around $z=1$~\cite{abramowitz+stegun}:
\begin{equation}
F_{\ell m}\propto~_2F_1\left(-\ell-s, \ell+1-s;1-s+2im\chi;1-z\right).
\end{equation}
Thus, the transformation for the Gaussian hypergeometric function as in Eq.~\eqref{transformaionfor2F1} implies that the asymptotic behavior of $F_{\ell m}$ at large distances takes the form of Eq.~\eqref{asymptoticphiellm:Kerr} ~\cite{abramowitz+stegun}:
\begin{equation}
\begin{split}
\label{ksell}
\kappa_{\ell m}:=&\frac{\Gamma(-2\ell-1)\Gamma(\ell+1-s)\Gamma(\ell+1+2im\chi)}{\Gamma(-\ell-s)\Gamma(2\ell+1)\Gamma(-\ell+2im\chi)},\\
=&-im\chi\frac{\left(-1\right)^{s}\left(\ell+s\right)!\left(\ell-s\right)!}{2\left(2\ell\right)!\left(2\ell+1\right)!}\prod_{j=1}^\ell\left[j^2\left(1-\chi^2\right)+m^2\chi^2\right],
\end{split}
\end{equation} 
which gives the analytic expression for the dissipation numbers, showing the vanishing of the spin-$s$-field Love numbers for all the multipoles~\cite{Chia:2020yla,Charalambous:2021mea,Hui:2022vbh}, i.e., ${\rm Re}[\kappa_{\ell m}^{(s)}]=0$.

\subsection{Remark: another perspective from an alternative equation}
We remark here the vanishing of the Love numbers of the Kerr black holes in terms of an alternative equation instead of the Teukolsky equation~\eqref{staticTeukolskyeq}. It is conjectured in Ref.~\cite{Hatsuda:2020iql} that the radial Teukolsky equation is isospectral, namely, having the same spectra as the following equation:
\begin{equation}
\label{Hatsudaeq}
\frac{\Delta}{z^2}\frac{d}{dz}\left(\frac{\Delta}{z^2}\frac{d\Phi_{\ell m}^{(s)}\left(z;\omega\right)}{dz}\right)+\left[\left(2M\omega\right)^2-V_{\ell m}^{(s)}(z)\right]\Phi_{\ell m}^{(s)}\left(z;\omega\right)=0,
\end{equation}
with
\begin{equation}
V_{\ell m}^{(s)}(z)=\frac{\Delta}{z^2}\left[4a^2\omega^2+\frac{4a\omega\left(m-a\omega\right)}{z}+\frac{_sA_{\ell m}+s(s+1)-a\omega\left(2m-a\omega\right)}{z^2}-\frac{s^2-1}{z^3}\right],
\end{equation}
where $M$ and $a$ are mass and spin parameter of the Kerr black hole, $\omega$ is a frequency of the linear fields, $_sA_{\ell m}$ is a separation constant of the spin-weighted spheroidal harmonics, and $\Delta$ and $z$ are given by Eqs.~\eqref{Delta:Kerr} and~\eqref{z:Kerr}, respectively. In Ref.~\cite{Hatsuda:2020iql}, it is suggested that the variable~$\Phi^{(s)}_{\ell m}$ is generated from the Newman-Penrose scalar by a highly-nontrivial integral transformation~\cite{Whiting:1988vc}.

Equation~\eqref{Hatsudaeq} can be reduced to the same form as the static Regge-Wheeler equation~\eqref{staticRWeq} smoothly in the static limit~$\omega\to 0$:
\begin{equation}
\label{staticHatsudaeq}
\frac{\Delta}{z^2}\frac{d}{dz}\left(\frac{\Delta}{z^2}\frac{d\Phi_{\ell m}^{(s)}}{dz}\right)-\frac{\Delta}{z^2}\left[\frac{\ell(\ell+1)}{z^2}-\frac{s^2-1}{z^3}\right]\Phi_{\ell m}^{(s)}=0.
\end{equation}
We can find the static limit leads to a surprising feature, i.e., the disappearance of $m$ in the master equation.
It should be noted, however, that for the validity of taking the static limit, the further investigation on the integral transformation in Ref.~\cite{Whiting:1988vc} is needed. 

If the conjecture above holds even for $\omega\to0$ and a physical boundary condition at the horizon~$z=1$ is $\Phi_{\ell m}^{(s)}|_{z\to1}\sim {\rm const}.$, the alternative equation~\eqref{staticHatsudaeq} may provide another symmetric perspective of the vanishing of the spin-$s$-field Love numbers of the Kerr black hole as follows: by introducing a new variable defined by $\phi_\ell^{(s)}(z)=\Phi_{\ell m}^{(s)}/z$ as in Eq.~\eqref{fieldredefinition}, one can show the existence of a hidden supersymmetric structure in a completely parallel manner as in the Schwarzschild black hole case and the variable~$\Phi_{\ell m}^{(s)}$ thus has no decaying terms at large distances under the requirement of $\Phi_{\ell m}^{(s)}|_{z\to1}\sim {\rm const}$.

\section{Summary and Discussion}
\label{Sec:Conclusion}
We have investigated the underlying symmetric structure leading to the vanishing of spin-$s$-field Love numbers of the Schwarzschild and Kerr black holes in terms of spacetime symmetry in a unified manner for the static perturbations. 
This is the first attempt to explain their no static response, i.e., vanishing of both Love and dissipation numbers, in a unified manner based on a symmetric approach from a geometrical point of view. We have also discussed the vanishing Love numbers  of the Kerr black hole with the nonzero dissipation numbers for the non-axisymmetric perturbations in terms of a radial constant found in a parallel manner as the axisymmetric field case.

The key observation is that the static spin-$s$ field perturbation to both the Schwarzschild and Kerr black holes can be reduced with the harmonic decomposition into a set of infinite static scalar fields in ${\rm AdS}_2$.
Here, the emergence of the
${\rm AdS}_2$ geometry is not derived from the enhancement of the isometry of the background spacetime itself. A slowly-varying perturbation to the Schwarzschild black hole can also be reduced into a set of infinite time-dependent scalar fields in ${\rm AdS}_2$ in a parallel manner.

The hidden supersymmetric structure exists for a static spin-$s$ field perturbation. In the reduced system of scalar fields in ${\rm AdS}_2$, each scalar field is associated with its pair, implying that all multipole modes of the perturbation can be regarded as symmetric partners which can be understood from the property of the supersymmetry algebra. The generator of the supersymmetric structure is constructed from a closed conformal Killing vector field of ${\rm AdS}_2$. Consequently, a radially conserved quantity exists. In particular, the no static response, i.e., the vanishing of both the spin-$s$-field Love and dissipation numbers of the Schwarzschild black hole, can be understood from this conserved quantity in a unified manner. This is also the case of the Kerr black hole for the axisymmetric perturbation.

In terms of the non-axisymmetric perturbation field to the Kerr black hole, the structure generated by the ladder operator is not prescribed under the regularity condition to the angular variables. Although its interpretation is controversial~\cite{Hui:2021vcv, Hui:2022vbh,Charalambous:2022rre}, we have shown the vanishing of the Love numbers with the nonzero dissipation numbers by using the radial constant found in a parallel manner as the axisymmetric field case. It should be emphasized that the aforementioned structure is nothing else but a reflection of the mathematical property of the current four-dimensional system even though the singular solutions and regular ones are connected by the ladder operator. We have left further discussions on~1)~the geometrical interpretation of the structure in the non-axisymmetric perturbation field on the Kerr background;~2)~whether or not the singular behavior is indeed problematic;~3)~the connection to Love symmetry~\cite{Charalambous:2022rre}, in outlook. 

The construction of the ladder operator is coordinate-independent. We then stress that the analysis with the ladder operator chose a specific coordinate but used a geometrically meaningful one: the conserved quantities are associated with the direction generated by the closed conformal Killing vector field. 

We comment on previous works that have independently studied the relation between the vanishing of Love numbers and hidden symmetries. While the generators of those hidden symmetries appear to come from geometrical~\cite{Hui:2021vcv,Hui:2022vbh} or algebraic properties~\cite{Charalambous:2021kcz,BenAchour:2022uqo,Charalambous:2022rre}, they would have deeper connection with the hidden effective ${\rm AdS}_2$ geometry. We particularly mention that our ladder operator includes 1) the generators of the ladder symmetry discussed in Ref.~\cite{Hui:2021vcv}, which arise from an isometry of the Euclidean ${\rm AdS}_3$ at least for scalar fields and are heuristically constructed for spin-$1,2$ fields, 2) the generator in given Ref.~\cite{BenAchour:2022uqo}, which is constructed in terms of the ${\rm SL}(2,\mathbb{R})$ symmetry. Our result supports the claim, ``the Love symmetry exactly reduces to the ${\rm SL}(2,\mathbb{R})$ isometry of the ${\rm AdS}_2$ near horizon black hole geometry'' in Ref.~\cite{Charalambous:2022rre}.

Our work can have several future extensions. First, it is interesting to analyze the vanishing of the Love numbers of the Kerr black hole in terms of the alternative equation~\eqref{Hatsudaeq} in Sec.~\ref{Sec:VanishingLoveofKerrBH}. Next, we expect that the vanishing of Love numbers of other kinds of black holes, e.g., the Schwarzschild black hole in the Brans-Dicke theory, the Reissner-Nordstr\"{o}m black hole in the Einstein-Maxwell theory~\cite{Cardoso:2017cfl,Cardoso:2019upw} etc., can also be understood in terms of another ``hidden'' symmetry. 
In our forthcoming paper, we show more than a few black holes with zero Love numbers exist either beyond General Relativity or in non-vacuum from hidden symmetries.
For testing theories of gravity by future gravitational-wave observations, we need deeper theoretical understanding on a system with or without non-zero Love numbers in advance.

\begin{acknowledgments}
The authors wish to express their cordial gratitude to Prof. Takahiro Tanaka, the Leader of Innovative Area Grants-in-Aid for Scientific Research ``Gravitational wave physics and astronomy: Genesis'', for his continuous interest and encouragement.
The authors also would like to thank Yasuyuki Hatsuda, Kazumi Kashiyama, Koutarou Kyutoku, Masato Nozawa, and Shijun Yoshida for fruitful discussions and useful comments.
This research is supported by Grants-in-Aid for Scientific Research (TK and KO: 17H06360, MK: 22K03626, HN: 17H06358, 21H01082, 21K03582, KO: 17H01102, 17H02869, 22H00149) from the Japan Society for the Promotion of Science. TK is also supported by VILLUM FONDEN (grant no. 37766), by the Danish Research Foundation, and under the European Union’s H2020 ERC Advanced Grant ``Black holes: gravitational engines of discovery'' grant agreement no. Gravitas–101052587. KO acknowledges support from the Amaldi Research Center funded by the MIUR program ``Dipartimento di Eccellenza'' (CUP:B81I18001170001). 

\end{acknowledgments}

\appendix

\section{Polar-type static tidal-field perturbation to Schwarzschild black holes: construction and tidal Love numbers}
\label{Appendix:symmetryofpolarperturbation}
Here, we show no static response for the polar-type perturbation with the aid of the Chandrasekhar transformation~\cite{Chandrasekhar:1975zza}. Static polar-type perturbations~$\Phi_\ell^+(z)$ are governed by the Zerilli equation in the static limit~\cite{PhysRevD.2.2141,Jhingan:2002kb}, 
\begin{equation}
\label{staticZerillieq}
\frac{\Delta}{z^2}\frac{d}{dz}\left(\frac{\Delta}{z^2}\frac{d\Phi_\ell^+}{dz}\right)-V_\ell^+(z)\Phi_\ell^+=0,
\end{equation}
where
\begin{equation}
V_\ell^+(z):=\frac{\Delta}{z^2}\left(\frac{9+9\lambda z+3\lambda^2z^2+\lambda^2(\lambda+2)z^3}{z^3(\lambda z+3 )^2}\right),
\end{equation}
with $\lambda=\ell^2+\ell-2$. Here, the index~$\ell=2,3,\cdots$ is that of multipole modes.   

\subsection{Construction of polar-type static tidal fields}
\label{ConstructionofpolartypeStaticTidalFields}
We generate polar-type perturbations~$\Phi_\ell^+$ from an axial-type perturbation~$\Phi_\ell^-(z)$ which satisfies the static Regge-Wheeler equation,
\begin{equation}
\label{staticRWeq:appendix}
\frac{\Delta}{z^2}\frac{d}{dz}\left(\frac{\Delta}{z^2}\frac{d\Phi_\ell^-}{dz}\right)-V_\ell^-(z)\Phi_\ell^-=0,
\end{equation}
with
\begin{equation}
V_\ell^-(z):=\frac{\Delta}{z^2}\left(\frac{\ell(\ell+1)}{z^2}-\frac{3}{z^3}\right),
\end{equation}
which is the same as Eq.~\eqref{staticRWeq:s=2}. We perform the Chandrasekhar transformation~\cite{Chandrasekhar:1975zza},
\begin{equation}
\label{isospectrality}
\Phi_\ell^-\to \tilde{\Phi}_\ell^+=D_+\Phi_\ell^-,
\end{equation}
where
\begin{equation}
\label{isoDplus}
D_+:=\frac{\Delta}{z^2}\frac{d}{dz}+\frac{\Delta}{z^3\left(1+\lambda z/3\right)}+\frac{\lambda(\lambda+2)}{6}.
\end{equation}
Then, one can show that $ \tilde{\Phi}_\ell^+$ satisfies the following equation,
\begin{equation}
\frac{\Delta}{z^2}\frac{d}{dz}\left(\frac{\Delta}{z^2}\frac{d \tilde{\Phi}_\ell^+}{dz}\right)-V_\ell^+ \tilde{\Phi}_\ell^+=0.
\end{equation}
This is nothing else but the static Zerilli equation~\eqref{staticZerillieq}. Thus, the function~$ \tilde{\Phi}_\ell^+$ describes a polar-type static tidal-field perturbation.

\subsection{No static response: quadrupole case~$\ell=2$}
We show the vanishing of the quadrupole tidal Love and dissipation numbers for the polar-type perturbation from the axial-type one. The general solution of the axial-type perturbation with $\ell=2$ is given in Eq.~\eqref{generalsol:l=2}, i.e., 
\begin{equation}
\begin{split}
\Phi_2^-=&C_{1}z^3+C_{2}\frac{3+4z+6 z^2+12 z^3+12z^4\ln\left(1-1/z\right)}{12z},
\end{split}
\end{equation}
where $C_{1}$, $C_{2}$ are constants. Acting the operator~$D_+$ in Eq.~\eqref{isoDplus}, we obtain
\begin{equation}
\begin{split}
\label{DPhiminusgeneral}
 \tilde{\Phi}_2^+=&C_{1}\frac{z\left(-3+6z^2+4z^3\right)}{3+4z}\\
&+C_{2}\frac{z\left(13+24z+12 z^2-12 z^3-3\left(3-6z^2-4z^3\right)\ln\left(1-1/z\right)\right)}{9+12z}.
\end{split}
\end{equation}
This is indeed a solution of the static Zerilli equation~\eqref{staticZerillieq} with $\ell=2$. Equation~\eqref{DPhiminusgeneral} shows that the term of $C_2$ is logarithmically divergent in the horizon limit~$z\to1$; to eliminate that term, we impose~$C_{2}=0$, thereby obtaining
\begin{equation}
\begin{split}
\label{DPhiminus}
 \tilde{\Phi}_2^+=&C_{1}\frac{z\left(-3+6z^2+4z^3\right)}{3+4z}.
\end{split}
\end{equation}
It follows that the solution~$ \tilde{\Phi}_2^+$ in Eq.~\eqref{DPhiminus} is regular at the horizon~$z=1$ and is purely growing at large distances~$z\gg1$, i.e., $\Phi_2^+|_{z\gg1}\sim z^3$. Thus, the quadrupolar tidal Love and dissipation numbers for the polar-type perturbation vanish. One can show that this is also the case of higher multipoles in the same manner.

\section{Reduction of a slowly-varying perturbation}
\label{Appendix:Reductionofalongwavelengthperturbation}
We reduce a slowly-varying spin-$s$ field perturbation into a time-dependent scalar field in ${\rm AdS_2}$, which satisfies the equation of motion with the same form as Eq.~\eqref{StaticscalarAdS2}.  This implies that the time-dependent spin-$s$ field also has  supersymmetric structure. Let us consider a time-dependent spin-$s$ field perturbation that satisfies the Regge-Wheeler equation~\cite{Regge:1957td}:
\begin{equation}
\label{RWeq}
-\partial_t^2\Phi_{\ell}\left(t,z\right)+\frac{\Delta}{z^2} \partial_z\left(\frac{\Delta}{z^2}\partial_z\Phi_\ell\left(t,z\right)\right)-\frac{\Delta}{z^2}\left(\frac{\ell(\ell+1)}{z^2}-\frac{s^2-1}{z^3}\right)\Phi_\ell\left(t,z\right)=0.
\end{equation}
Now, we assume that the time dependence of the perturbation is weak: the time scale of the change of the field is much longer than the black hole radius, i.e., $|\partial_t\Phi_\ell|^{-1}\gg1$. We further focus on a ``near zone'' such that the time dependence of the field is still weak at a distant region from the black hole horizon, i.e., $1<z\ll |\partial_t\Phi_\ell|^{-1}$. We then have
\begin{equation}
\begin{split}
\frac{z^4}{\Delta}\partial_t^2\Phi_\ell=&\frac{1}{\Delta}\partial_t^2\Phi_\ell+z^2\partial_t^2\Phi_\ell+z\partial_t^2\Phi_\ell+\partial_t^2\Phi_\ell+\frac{1}{z}\partial_t^2\Phi_\ell\\
\simeq& \frac{1}{\Delta}\partial_t^2\Phi_\ell,
\end{split}
\end{equation}
where $\simeq$ means an equality within the approximations~$z\ll |\partial_t\Phi_\ell|^{-1}$ and $|\partial_t\Phi_\ell|^{-1}\gg1$. This corresponds to the near-zone approximation~\cite{Castro:2010fd,Bertini:2011ga}. 
Using this property, we obtain an equation for the slowly-varying spin-$s$ field perturbation in the near zone~$1<z\ll |\partial_t\Phi_\ell|^{-1}$ from Eq.~\eqref{RWeq}:
\begin{equation}
\label{effectiveRWeq}
-\frac{1}{\Delta}\partial_t^2\Phi_{\ell}+z^2 \partial_z\left(\frac{\Delta}{z^2}\partial_z\Phi_\ell\right)-\left(\ell(\ell+1)-\frac{s^2-1}{z}\right)\Phi_\ell=0.
\end{equation}

We redefine the variable~$\Phi_\ell$ in Eq.~\eqref{effectiveRWeq} as
\begin{equation}
\label{fieldredefinition:timedependent}
\phi_\ell(t,z)=\frac{\Phi_{\ell}}{z}.
\end{equation}
One can then obtain
\begin{equation}
\label{EOMAdS2:timedependent}
\left[\Box_{{\rm AdS}_2}-\left(\ell(\ell+1)-\frac{s^2}{z}\right)\right]\phi_\ell=0.
\end{equation}
Here, the d'Alembertian~$\Box_{{\rm AdS}_2}$ is given by
\begin{equation}
\square_{{\rm AdS}_2}=-\frac{1}{\Delta}\partial_t^2+\partial_z\left(\Delta\partial_z\right),
\end{equation}
which is the same as Eq.~\eqref{BoxAdS2general}, and is based on the line element of ${\rm AdS}_2$,
\begin{equation}
g_{ab}^{{\rm AdS}_2}dx^a dx^b
=-\Delta dt^2+\frac{1}{\Delta}dz^2,
\end{equation}
which is the same as Eq.~\eqref{AdS2metric}. Thus, the slowly-varying spin-$s$ field perturbation can be reduced to a set of infinite  time-dependent scalar fields in ${\rm AdS}_2$. As can be seen from the fact that Eq.~\eqref{EOMAdS2:timedependent} is the same form as Eq.~\eqref{StaticscalarAdS2}, one can define the ladder operator~\eqref{LOs} and can show a hidden symmetric structure in a completely parallel manner as the static-field case.

\section{Ladder operators from spacetime conformal symmetry and application to perturbations to Schwarzschild-Tangherlini black holes}
\label{Appendix:ConditionforVerticalSymmetry}
We here derive a ladder operator from spacetime conformal symmetry of ${\rm AdS}_2$. In particular, we show that the systems~\eqref{StaticscalarAdS2} and~\eqref{staticfieldinAdS2:Kerr} naturally appear under the requirement of the existence of a ladder operator for a scalar field in general systems. We further discuss the application to a slowly-varying scalar field perturbation to the Schwarzschild-Tangherlini black holes~\cite{Tangherlini:1963bw}.

\subsection{Generic time-dependent scalar field in ${\rm AdS}_2$}
We consider a time-dependent scalar field in ${\rm AdS}_2$, which satisfies
\begin{equation}
\label{EoMforPsinu}
\left[\Box_{{\rm AdS}_2}-\left( {\hat{\ell}}({\hat{\ell}}+1)+P\right)\right]\Psi_{\hat{\ell}}=0,
\end{equation}
where ${\hat{\ell}}$ is a nonnegative real-valued constant. Note that $\hat{\ell}$ is not necessarily an integer. Here, we have defined the d'Alembertian in ${\rm AdS_2}$, $\Box_{{\rm AdS}_2}$,  which corresponds to Eq.~\eqref{BoxAdS2general} in the $(t,z)$ coordinates. For later convenience, we introduce a new real-valued parameter~$k$ such that
\begin{equation}
k(k+1)={\hat{\ell}}({\hat{\ell}}+1).
\end{equation}
Solving this quadratic equation for $k$, we obtain
\begin{equation}
\label{kpmell}
k_+=-\hat{\ell}-1,~~k_-=\hat{\ell},
\end{equation}
where we have assigned $k_\pm$ on the solutions so that $k_+<k_-$. Note that $k_{\pm}$ is an integer if and only if $\hat{\ell}$ is an integer. In terms of $k_\pm$, Eq.~\eqref{EoMforPsinu} can be rewritten as
\begin{equation}
\label{EoMforPsi}
\left[\Box_{{\rm AdS}_2}-\left( k_\pm(k_\pm+1)+P\right)\right]\Psi_{\hat{\ell}}=0.
\end{equation}

\subsection{Conditions for commutation relations holding}
We first require a commutation relation:
\begin{equation}
\label{commutationrelation:general}
\left[\Box_{{\rm AdS_2}}-P,\mathcal{D}_{k_\pm}\right]=-2 k_\pm\mathcal{D}_{k_\pm}+2Q\left[\Box_{{\rm AdS}_2}-\left( k_\pm(k_\pm+1)+P\right)\right],
\end{equation}
where $\mathcal{D}_{k_\pm}$ is a derivative operator, and $Q$ is a function in ${\rm AdS}_2$. Acting this commutation relation on a smooth function~$\Psi_{\hat{\ell}}$ leads to
\begin{equation}
\left[\Box_{{\rm AdS}_2}-\left( \left(k_\pm-1\right)k_\pm+P\right)\right]\mathcal{D}_{k_\pm}\Psi_{\hat{\ell}}=\left(\mathcal{D}_{k_\pm}+2Q\right)\left[\Box_{{\rm AdS}_2}-\left( k_\pm(k_\pm+1)+P\right)\right]\Psi_{\hat{\ell}}.
\end{equation}
One can see that when $\Psi_{\hat{\ell}}$ is a solution of Eq.~\eqref{EoMforPsi}, the right-hand side vanishes, obtaining
\begin{equation}
\label{Comuk}
\left[\Box_{{\rm AdS}_2}-\left( \left(k_\pm-1\right)k_\pm+P\right)\right]\mathcal{D}_{k_\pm}\Psi_{\hat{\ell}}=0.
\end{equation}
Noting that $(k_+-1)k_+=(\hat{\ell}+1)(\hat{\ell}+2)$ and $(k_--1)k_-=(\hat{\ell}-1)\hat{\ell}$, this is an equation of motion for a scalar field~$\mathcal{D}_{k_\pm}\Psi_{\hat{\ell}}$ with a shifted parameter $\hat{\ell}\to \hat{\ell}\pm1$, implying that the operator~$\mathcal{D}_{k_\pm}$ generates another scalar field~$\mathcal{D}_{k_\pm}\Psi_{\hat{\ell}}$ with $\hat{\ell}\to \hat{\ell}\pm1$ keeping the potential function~$P$ unchanged.

Now, we seek for ladder operators~$\mathcal{D}_{k_\pm}$ and potentials~$P$ such that the commutation relation~\eqref{commutationrelation:general} holds. We begin with introducing a form of a first-order differential operator,
\begin{equation}
\mathcal{D}_{k_\pm}=V^a\nabla_a+\mathcal{K},
\end{equation}
where $V^a$ and ${\cal K}$ are a vector field and a function in ${\rm AdS}_2$, respectively. Then, the left-hand side of the commutation relation~\eqref{commutationrelation:general} is calculated to
\begin{equation}
\left[\Box_{{\rm AdS_2}}-P,\mathcal{D}_{k_\pm}\right]=2\left(\nabla^a V^b\right)\nabla_a\nabla_b+\left(\Box_{{\rm AdS}_2}V^b+V_aR^{ab}+2\nabla^b{\cal K}\right)\nabla_b+\Box_{{\rm AdS}_2}{\cal K}+V^a\nabla_aP,
\end{equation}
where $R_{ab}$ is the Ricci tensor of ${\rm AdS}_2$. Here, we have assumed that the commutation relation acts on a smooth function. The right-hand side is calculated to
\begin{equation}
\begin{split}
&-2 k_\pm \mathcal{D}_{k_\pm}+2Q\left[\Box_{{\rm AdS}_2}-\left( k_\pm(k_\pm+1)+P\right)\right]\\
&=2Q\Box_{{\rm AdS}_2}-2 k_\pm V^a\nabla_a+\left[-2 k_\pm{\cal K}-2Q\left( k_\pm(k_\pm+1)+P\right)\right].
\end{split}
\end{equation}
With these explicit forms, the commutation relation~\eqref{commutationrelation:general} is divided into the following three conditions:
\begin{equation}
\label{CKVeq}
\nabla_a V_b+\nabla_b V_a=2Qg_{ab},
\end{equation}
\begin{equation}
\label{secondeq}
\Box_{{\rm AdS}_2}V^a+R^{ab}V_b+2\nabla^a{\cal K}=-2 k_\pm V^a,
\end{equation}
\begin{equation}
\label{thirdeq}
\Box_{{\rm AdS}_2}{\cal K}+V^a\nabla_aP=-2 k_\pm {\cal K}-2Q\left[k_\pm (k_\pm+1)+P\right].
\end{equation}

Let us investigate these conditions one by one. We focus on the case~$Q\neq 0$.\footnote{For the case~$Q=0$, Eq.~\eqref{CKVeq} corresponds to a Killing equation. All the Killing vector fields are not compatible with the conditions~\eqref{secondeq} and~\eqref{thirdeq} for the time-independent potential~$P(z)$.} 
Then, Eq.~\eqref{CKVeq} corresponds to a conformal Killing equation. Taking the trace of Eq.~\eqref{CKVeq}, we obtain
\begin{equation}
Q=\frac{1}{2}\nabla_aV^a.
\end{equation}
For the conformal Killing vector field~$V^a$ in two dimensions, we have
\begin{equation}
\label{KillingCondition}
\Box_{{\rm AdS}_2}V_a+R_{ab}V^b=0.
\end{equation}
Therefore, Eq.~\eqref{secondeq} is reduced to
\begin{equation}
\label{secondsecondeq}
\nabla^a{\cal K}=- k_\pm V^a.
\end{equation}
Acting $\epsilon^{ab}\nabla_a$ with the Levi-Civita tensor~$\epsilon^{ab}$ from the left-hand side leads to 
\begin{equation}
\epsilon^{ab}\nabla_a\nabla_b{\cal K}=- k_\pm \epsilon^{ab}\nabla_aV_b.
\end{equation}
The left-hand side vanishes, thereby yielding a condition,
\begin{equation}
\nabla_aV_b=\nabla_bV_a.
\end{equation}
This means that the conformal Killing vector fields~$V^a$ are ``closed''. In ${\rm AdS}_2$, there exist three independent closed conformal Killing vector fields. Then, the conditions~\eqref{CKVeq}--\eqref{thirdeq} reduce to
\begin{equation}
\label{CCKVeq}
\nabla_a V_b=\frac{1}{2}\left(\nabla_c V^c\right)g_{ab},
\end{equation}
which is the closed conformal Killing equation, and
\begin{equation}
\label{resecondeq}
\nabla^a{\cal K}=-k_\pm V^a,
\end{equation}
which is the same as Eq.~\eqref{secondsecondeq}, and 
\begin{equation}
\label{rethirdeq}
\Box_{{\rm AdS}_2}{\cal K}+V^a\nabla_aP=-2 k_\pm {\cal K}-\left(\nabla_a V^a\right)\left[ k_\pm (k_\pm+1)+P\right].
\end{equation}
Because $V^a = \nabla^a Q$ in ${\rm AdS}_2$,
Eq.~\eqref{resecondeq} can be solved as
\begin{equation}
\label{deneralK}
{\cal K} = -k_\pm Q - \frac{c_0}{2k_\pm},
\end{equation}
where $c_0$ is a constant. We thus obtain the general form of the ladder operator,
\begin{equation}
\label{generalD}
{\cal D}_{k_\pm}=\mathcal{L}_{V} -\frac{k_\pm}{2}\left(2 Q + \frac{c_0}{k_\pm^2}\right),
\end{equation}
where the first term of the right-hand side is the Lie derivative with respect to the closed conformal Killing vector field. We also obtain the general form of $P$ from Eq.~\eqref{rethirdeq},
\begin{equation}
\label{generalP}
P = \frac{c_0 Q+ c_p}{V_a V^a},
\end{equation}
where $c_p$ is a constant.
Note that $c_p$ can depend on the coordinate value which is perpendicular to 
the integral curve of $V^a$. Equation~\eqref{generalP} shows that the functional form of the potential $P$ is strongly restricted, 
i.e., $P$ depends on the closed conformal Killing vector field~$V^a$ and two parameters~$c_0$ and $c_p$.

\subsection{Ladder operators from spacetime conformal symmetry}
We here introduce the coordinate system~$(t,z)$; then, the line element of ${\rm AdS_2}$ is described by
\begin{equation}
\label{generalAdS2metric}
g_{ab}^{{\rm AdS}_2}dx^a dx^b=-\Delta dt^2+\frac{1}{\Delta}dz^2,~~\Delta=z(z-1),
\end{equation}
which is the same as Eq.~\eqref{AdS2metric}. Hereafter, we assume that $P$ is a function of $z$. From the conditions~\eqref{CCKVeq}--\eqref{rethirdeq}, we derive ladder operators and possible potential forms. Although there exist three independent closed conformal Killing vector fields satisfying Eq.~\eqref{CCKVeq}, we use here solely:
\begin{equation}
\begin{split}
\label{CCKV0}
V^a\frac{\partial}{\partial x^a}=&z(z-1) \frac{\partial}{\partial{z}}.
\end{split}
\end{equation}
Note that the assumption of the time independence of the potential forces one to choose the closed conformal Killing vector field~\eqref{CCKV0} only. The other closed conformal Killing vector fields are not compatible with the conditions~\eqref{resecondeq} and~\eqref{rethirdeq}; however, if we admit a time dependence of the potential, a ladder operator can be constructed in the almost parallel manner as in the present subsection.

In the current coordinate system, ${\cal K}$ in Eq.~\eqref{deneralK} takes the form,
\begin{equation}
\label{functionK}
{\cal K} = -\frac{k_\pm}{2}\left( 2z-1 +\frac{c_0}{k_\pm^2}\right),
\end{equation}
which leads to the operator,
\begin{equation}
\label{LO:appendix}
\mathcal{D}_{k_\pm}=z(z-1) \partial_{z}-\frac{k_\pm}{2}\left( 2z-1 + \frac{c_0}{k_\pm^2}\right).
\end{equation}
Equation~\eqref{generalP} leads to
\begin{equation}
P=\frac{ c_0\left(2z-1\right)+2c_p}{2z(z-1)}.
\end{equation}
Thus, the ladder operator~\eqref{LO:appendix} exists in the system,
\begin{equation}
\label{system:general}
\left[\Box_{{\rm AdS}_2}-\left(k_{\pm}(k_{\pm}+1)+\frac{ c_0\left(2z-1\right)+2c_p}{2z(z-1)}\right)\right]\Phi_{\hat{\ell}}=0.
\end{equation}

If we assign $c_0=-c_p$, the effective potential is regularized at $z=1$, leading to
\begin{equation}
\label{ScalarinAdS2withLadderOperator}
\left[\Box_{{\rm AdS}_2}-\left(k_{\pm}(k_{\pm}+1)-\frac{c_p}{z}\right)\right]\Phi_{\hat{\ell}}=0,
\end{equation}
which admits the ladder operator,
\begin{equation}
\label{finalD}
\mathcal{D}_{k_\pm}=z\left(z-1\right) \partial_{z}-\frac{ k_\pm}{2}\left(2z-1-\frac{c_p}{ k_\pm^2}\right).
\end{equation}
In particular, if $k_\pm$ in Eq.~\eqref{kpmell} is defined from $\hat{\ell}=\ell~(\ell=s,s+1,s+2\cdots; s=0,1,2)$, and $c_p=s^2$, the system~\eqref{ScalarinAdS2withLadderOperator} and the operator~\eqref{finalD} correspond to Eqs.~\eqref{StaticscalarAdS2} and~\eqref{laderoperator}, respectively.

If we choose $c_0=-2im\chi s$, $c_p=-(4m^2\chi-s^2)/4$, and $\hat{\ell}=\ell~(\ell=s,s+1,s+2,\cdots)$, the system~\eqref{system:general} becomes
\begin{equation}
\left[\Box_{{\rm AdS}_2}-\left(k_\pm(k_\pm+1)-\frac{4m^2\chi^2-s^2}{4z(z-1)}+im\chi s \frac{1-2z}{z(z-1)}\right)\right]\Phi_{\hat{\ell}}=0,
\end{equation}
which is the same as Eq.~\eqref{staticfieldinAdS2:Kerr}. We also have the ladder operator,
\begin{equation}
\mathcal{D}_{k_\pm}=z(z-1)\frac{d}{dz}-\frac{k_\pm}{2}\left(2z-1-i\frac{2 m\chi  s}{k_\pm^2}\right),
\end{equation}
which corresponds to Eq.~\eqref{LOs:Kerr}.

\subsection{Application: perturbations to the Schwarzschild-Tangherlini black hole}
We discuss the application of the above ladder operators to a problem of a slowly-varying perturbation to the Schwarzschild-Tangherlini black hole in $(n+2)$ dimensions, which is described by~\cite{Tangherlini:1963bw}
\begin{equation}
ds^2=-\left(1-\frac{1}{\hat{z}^{n-1}}\right)dt^2+\frac{1}{1-\frac{1}{\hat{z}^{n-1}}}d\hat{z}^2+\hat{z}^2d\Omega_{n}^2,
\end{equation}
where $\hat{z}$ is a dimensionless areal coordinate and $d\Omega_{n}^2$ is the line element of the unit sphere in $n$ dimensions. 

Let us first consider the four-dimensional case~$n=2$, i.e., the Schwarzschild black hole case. As seen in Appendix~\ref{Appendix:Reductionofalongwavelengthperturbation}~(see also Sec.~\ref{Sec:HiddenSymmetryfromAdS2}), the slowly-varying spin-$s$ field perturbation can be reduced to a set of infinite time-dependent scalar fields satisfying an equation~(see Eqs.~\eqref{EOMAdS2:timedependent} and~\eqref{StaticscalarAdS2}), which take the form of Eq.~\eqref{ScalarinAdS2withLadderOperator} with $c_p=s^2$, $z=\hat{z}$, and $\hat{\ell}=\ell$. Then, the ladder operators in Eq.~\eqref{finalD} with $c_p=s^2$ shift the multipole index~$\ell$ into $\ell\pm1$.

Let us next consider the higher-dimensional case~$n\ge3$. A slowly-varying scalar field perturbation can be reduced to a set of infinite time-dependent scalar fields satisfying an equation that takes the form of Eq.~\eqref{ScalarinAdS2withLadderOperator} with $c_p=0$, $z=\hat{z}^{n-1}$, and ${\hat{\ell}}=\ell/(n-1)$~(cf., discussion around Eq.~(4.48) in Ref.~\cite{Charalambous:2022rre}). Then, the ladder operators~\eqref{finalD} with $c_p=0$ shift the multipole index~${\hat{\ell}}$ into ${\hat{\ell}}\pm1$. Thus, in the higher dimensions, the ladder operator relates a mode with a given $\ell$ with not all modes but with another mode with~$\ell\propto n-1$ for the Schwarzschild-Tangherlini black hole.

\section{No static response for polar-type perturbations from the hidden supersymmetric structure}
\label{Appendix:tidalLovenumberofPolar}
Here, we show no static response for the polar-type tidal field perturbation in terms of the radially conserved quantity associated with a symmetric structure. We map the polar-type perturbation into the axial-type one with the Chandrasekhar transformation~\cite{Chandrasekhar:1975zza} and then analyze the problem by following the strategy in Sec.~\ref{Sec:VanishingLovefromSymmetry}.

\subsection{Construction of axial-type static tidal fields}
\label{ConstructionofaxialtypeStaticTidalFields}
In the parallel manner as in Appendix~\ref{Appendix:symmetryofpolarperturbation}, we generate axial-type perturbations~$\Phi_\ell^-$ from a polar-type perturbation~$\Phi_\ell^+$ with the Chandrasekhar transformation~\cite{Chandrasekhar:1975zza},
\begin{equation}
\label{isospectralityminus}
\Phi_\ell^+\to \tilde{\Phi}_\ell^-=D_{-}\Phi_\ell^+,
\end{equation}
where
\begin{equation}
\label{isoDminus}
D_{-}:=\frac{\Delta}{z^2}\frac{d}{dz}-\frac{\Delta}{z^3\left(1+\lambda z/3\right)}-\frac{\lambda(\lambda+2)}{6},
\end{equation}
with $\lambda=\ell^2+\ell-2$. One can then show that the function~$ \tilde{\Phi}_\ell^-$ satisfies the static Regge-Wheeler equation~\eqref{staticRWeq:appendix}. Introducing a new variable,
\begin{equation}
\tilde{\phi_\ell}(z):= \frac{\tilde{\Phi}_\ell^-}{z},
\end{equation}
the static Regge-Wheeler equation for $ \tilde{\Phi}_\ell^-$ leads to
\begin{align}
\left[\square_{{\rm AdS}_2}-\left(\ell(\ell+1)-\frac{4}{z}\right)\right]\tilde{\phi_\ell}=&0.
\end{align}
This is the same form as Eq.~\eqref{StaticscalarAdS2}. 

\subsection{No static response from the hidden supersymmetric structure}
We construct a radially conserved quantity for $\tilde{\phi_\ell}$ in the same manner as in Sec.~\ref{Sec:VanishingLovefromSymmetry}:
\begin{equation}
\label{ConservedQellplus}
Y_\ell^+:= z^{2}\mathcal{D}_{3} \cdots\mathcal{D}_{\ell-1}\mathcal{D}_{\ell} \tilde{\phi_\ell},
\end{equation}
which satisfies
\begin{equation}
\label{conservationlawfortildephiell}
\frac{d}{dz}Y_\ell^+=0.
\end{equation}
In the same manner as in Sec.~\ref{Sec:VanishingLovefromSymmetry}, one can show that the regular solution at the horizon connects to the purely growing solution at large distances in terms of the radially conserved quantity~$Y_\ell^{+}$, showing the vanishing of the tidal Love and dissipation numbers for the polar-type tidal field perturbation for all $\ell$.

\bibliographystyle{unsrt}
\bibliography{refs}

\end{document}